%% file: mblithium.tex
\documentclass[fleqn,usenatbib]{mnras}

\usepackage{newtxtext,newtxmath}

\usepackage[T1]{fontenc}
\usepackage{ae,aecompl}

\usepackage{graphicx}	
\usepackage{amsmath}	
\usepackage{amssymb}	
\usepackage{nccmath}
\usepackage{isotope}
\usepackage{float}
\usepackage{caption}
\usepackage{subcaption}
\usepackage{multicol}
\usepackage[normalem]{ulem}

\input definitions.tex

\title[Rotation, magnetic braking \& Li abundances]{Study of the effects of magnetic braking on the lithium abundances of the Sun and solar-type stars}

\author[R. Caballero Navarro et al.]{
R. Caballero Navarro,$^{1}$\thanks{E-mail: rcaballeron@correo.ugr.es}
A. Garc\'ia Hern\'andez,$^{1,2}$\thanks{E-mail: agh@ugr.es}
A. Ayala$^{2},$\thanks{E-mail: aayala@ugr.es}
J.~C. Su\'arez$^{1,2}$\thanks{E-mail: jcsuarez@ugr.es}
\\
$^{1}$Dept. Theoretical Physics and Cosmology, University of Granada (UGR), 18071, Granada, Spain\\
$^{2}$Instituto de Astrof\'isica de Andaluc\'ia (CSIC), Glorieta de la Astronom\'ia S/N, 18008, Granada, Spain\\
}

\date{Accepted 2020 May 29. Received 2020 April 17; in original form 2019 November 30}

\pubyear{2019}

\begin{document}
\label{firstpage}
\pagerange{\pageref{firstpage}--\pageref{lastpage}}
\maketitle

\begin{abstract}
The study of lithium (Li) surface abundance in the Sun and young stellar globular clusters which are seemingly anomalous in present-day scenarios, as well as the influence of rotation and magnetic braking (MB) on its depletion during pre-main sequence (PMS) and main sequence (MS).
In this work, the effects of rotational mixing and of the rotational hydrostatic effects on Li abundances are studied by simulating several grids of PMS and MS rotating and non-rotating models. Those effects are combined with the additional impact of the MB (with magnetic field intensities ranging between 3.0 and 5.0 G). The data obtained from simulations are confronted by comparing different stellar parameters. The results show that the surface Li abundance for the Sun like models at the end of the PMS and throughout the MS decreases when rotational effects are included, i.e. the Li depletion rate for rotating models is higher than for non-rotating ones. This effect is attenuated when the MB produced by a magnetic field is present. This physical phenomena impacts also the star effective temperature ($\teff$) and its location in the HR diagram. The impact of MB in Li depletion is sensitive to the magnetic field intensity: the higher it is, the lower the Li destruction. A direct link between the magnetic fields and the convective zone (CZ) size is observed: stronger magnetic fields produce shallower CZ's. This result suggests that MB effect must be taken into consideration during PMS if we aim to reproduce Li abundances in young clusters.
\end{abstract}




\section{Introduction} \label{sec_intro}
Despite decades of theoretical efforts, a coherent explanation for the Li abundance discrepancies detected in star clusters cannot be found. In young and old clusters these differences have been documented for stars in both PMS and MS  evolutionary stages. Additionally, these theoretical models are not able to explain the abundances detected in the late stages of the MS \citep{Tschape2001}.\par

Standard models, models that include convection only as a mixing process and do not consider other transport options like diffusion or angular momentum loss (AML), have been mainly involved in the elaboration of these predictions \citep{Sestito2005}. Li is destroyed at a temperature $\tli \approx 2.5 x 10^6\; K$ when a Li atom collides with a proton producing two He atoms. That takes place during proton-proton type II reactions (P-P II), and is therefore directly destroyed in stellar envelopes when the temperature at the base of the convection zone (BCZ) reaches  $\tli$. The Sun in particular and the solar-type stars in general are characterized by having a CZ that covers much of the stellar radius during the PMS which causes that their lower limit to exceed $\tli$ \citep{Iben1965}. This depletion stops during the approach to the zero-age MS (ZAMS) when the convection zone retreats and the temperature at BCZ is cooler than $\tli$. In the standard models only the mass and the initial chemical composition determine at what distance from the stellar surface $\tli$ temperature is reached. Therefore stars with similar mass belonging clusters are expected to reach the ZAMS with equal surface Li abundances. Furthermore, they should also show a very similar Li evolution until their approach to the terminal-age MS (TAMS). During this period the convection mechanisms unleash a mixing process that homogenizes the chemical composition of the envelope, from its lower limit till the star surface. However, different abundances of Li have been observed for different stellar populations \citep[see][and references therein]{Somers2014}.\par

Theoretical models take initial Li abundance as an input parameter and only describe how it is exhausted. Therefore, in order to make an accurate estimate of the initial abundance of Li, it is a prerequisite to be able to compare observations and models beforehand. Our Sun represents a unique exception, since it allows us to know the current abundance of this element in its photosphere, $A(\isotope[7]{Li}) = 1.1 \pm 0.1 \, dex$ \citep{Jeffries2004}, where $A(\isotope[7]{Li})$ is defined

\begin{ceqn}
\begin{align}
    A(\isotope[7]{Li}) &= log(N_{\isotope[7]{Li}} / N_{\isotope[1]{H}}) + 12
    \label{eq:A_Li}
\end{align}
\end{ceqn}

and its initial abundance of $A(\isotope[7]{Li}) = 3.34 \, dex$ is obtained from meteorite measurements \citep{Randich2006}. For other stars the initial Li abundance is found to be in the interval $3.0 \, dex < A(\isotope[7]{Li}) < 3.4 \, dex$. This has been estimated from  \citet{Randich2006}.\par

The impact of rotation both on PMS and Li depletion for solar-type stars has already extensively debated in the past \citep{Pinsonneault1997,Jeffries2004,Somers2014} and revised more recently on the basis of the availability of more accurate measures \citep{Bouvier2016}. However, these previous studies focused mainly on the hydrostatic effects and did not consider the influence of a coupling between the stellar magnetic field and its possible spin-down effect. AML has a direct influence in the mixing processes. It can be produced by a number of relevant mechanisms: mass loss, magnetic fields and gravity waves (g-modes). Gravity waves \citep{Charbonnel2005} and magnetic fields \citep{Eggenberger2009} transmit angular momentum (AM) much more effectively than inducing mixing \citep{Denissenkov2007}. As a consequence, the amount of differential rotation between the radiative and the convective zones of the star is reduced. Additionally, magnetic fields would also originate in the limits between these zones, in the so-called tachoclines \citep{Aschwanden2014, Guerrero2016}, which interact with the solar wind particles that are forced to co-rotate with it. This slows down the star, effect which is known as magnetic braking (MB). \par

Today, standard models are not able to replicate the observed values of Li abundance on stellar surfaces. This might happen because certain physical mechanisms that influence the destruction of Li are either being modelled improperly or simply not being taken into account, e.g. the magnetic braking. In this paper we study the interactions between rotation and magnetic fields when it comes to discuss the AM distribution. \par

\section{Method} \label{sec_method}

In the remainder of this paper, we describe a semi-empirical approach for the calculation of the AML as a result of the torque applied by a magnetically-coupled stellar winds. This has been implemented as an extension to Modules for Experiments in Stellar Astrophysics stellar evolution code \citep[MESA; ][]{Paxton2011, Paxton2013,Paxton2015, Paxton2018, Paxton2019}.

\subsection{Modelling magnetic braking} \label{mod_mb}
We start by enumerating the most relevant aspects and assumptions made in modelling the evolution of rotation, magnetic braking and angular momentum in MESA. These will be considered for the calculation of the AML as a result of the torque applied by a magnetically-coupled stellar winds.\par 

A relevant parameter to characterize the influence of a given magnetic field on the stellar wind is the denominated wind confinement magnetic parameter ($\eta_*$). It represents a energy ratio (Eq.~\ref{eq:wind_conf}) and defines a characteristic parameter for the relative effectiveness of the magnetic fields in circumscribing and/or channeling the wind outflow \citep{UdDoula2002}.\par

\begin{ceqn}
\begin{equation}
    \eta_* = \frac{B^{2}/8\upi}{\rho\nu^2/2} \label{eq:wind_conf}
\end{equation}
\end{ceqn}

where $B^{2}/8\upi$ is the magnetic field energy density, $\frac{1}{2}\rho\nu^{2}$ the kinetic energy density, $B$ the magnetic field intensity, $\rho$ the mass density, and $\nu$ the stellar wind velocity.\par

Stars with similar initial masses but different mass loss ($\Dot{M}$) ratios will end up evolving very differently. The ionized particles carried by the solar wind not only contribute to the mass loss but also to the loss of kinetic ($K_e$) energy that is deposited in the interstellar medium. Given a star with a spherically symmetric wind, $\Dot{M}$ is characterized by the following expression:

\begin{ceqn}
\begin{equation}
    \Dot{M} = 4\upi r^2\rho\nu \label{eq:mass_loss}
\end{equation}
\end{ceqn}

Using (\ref{eq:wind_conf}) and (\ref{eq:mass_loss}), $\eta_*$ can be approximated by: 
\begin{ceqn}
\begin{equation}
    \eta_* = \frac{B^{2}r^{2}}{\Dot{M}\nu} \label{eq:wind_conf2}
\end{equation}
\end{ceqn}

In general, a magnetically channeled outflow has a complex stream geometry but for convenience, the Eq.~\ref{eq:mass_loss} simply characterizes the wind strength in terms of a spherically symmetric mass-loss rate. $\nu$ can be characterized by the radial variation of outflow velocity in terms of the velocity law (Eq.~\ref{eq:vel_law})
\begin{ceqn}
\begin{equation}
    \nu(r) = \nu_\infty (1-\rstar/r) \label{eq:vel_law}
\end{equation}
\end{ceqn}
where $r$ represents the distance from the center of the star to the point where the velocity is to be measured, $R_*$ is the radius of the star, and $\nu_\infty$ is the terminal wind velocity. $\nu_\infty$ is defined as the velocity that the wind reaches at large distance from the central star, where it is not accelerated anymore by the wind driving force but its deceleration due to interaction with the interstellar medium (ISM) is negligible \citep{Niedzielski2002}.\par

The line-driven winds of OB stars have terminal velocities that scale with the photospheric escape velocity \citep{Lamers2000} according to:
\begin{ceqn}
\begin{align}
    \nu_\infty &\simeq 1.92 \;\nu_{{\rm esc}} \label{eq:vinf}\\
    \nu_{{\rm esc}} &= \sqrt{\frac{2\,G\,\mstar}{\rstar}} \label{eq:vesc}
\end{align}
\end{ceqn}
where Eq.~\ref{eq:vesc} describes the Newtonian escape velocity from the stellar surface, where $G$ is the gravitational constant,   ${\rstar}$ the radius, and ${\mstar}$ the mass of the star.

Making use of (\ref{eq:vesc}), $\nu_\infty$ can be reformulated as:

\begin{ceqn}
\begin{equation}
    \nu_\infty \simeq 1.92 \; x \; 618 \; \Bigg(\sqrt{\frac{\rsun}{\rstar}\frac{\mstar}{\msun}} \;\Bigg) \label{eq:vinf}
\end{equation}
\end{ceqn}
where ${\rsun}$ is the radius, and ${\msun}$ the mass of the Sun.

It has been observed that strongly magnetic intermediate-mass stars typically have rotation rates much slower than other stars in their parent population \citep{Mathys2006}. In those stars, the magnetic fields interacts with the mass loss, where the Alfv\'{e}n radius ($R_{A}$) plays an important role. Making use of $\eta_*$ (Eq. \ref{eq:wind_conf}), $R_{A}$ is defined as the point in which the magnetic field energy density and the kinetic energy density are balanced. In case that $R_{A}$ is greater than the stellar radius, then the wind flow will have to follow the magnetic field. As a consequence, the material leaves the stellar surface with a higher specific AM, as the co-rotation radius has increased and it roughly corresponds to $R_{A}$. Following \citet{Weber1967} the AML can be calculated by:
\begin{ceqn}
\begin{equation}
 \Dot{J} = \frac{2}{3} \Dot{M}\Omega R^{2}_{A} \label{eq:j_dot}
\end{equation}
\end{ceqn}
where $\Dot{M}$ is the mass loss rate, $\Omega$ the angular velocity at star surface and $R_A$ the Alfv\'{e}n radius. \par

The above expression can be rewritten according to \citet{Ud-Doula2008} so that it depends on the $\eta_*$ instead of the $R_A$ as follows:
\begin{ceqn}
\begin{equation}
 \Dot{J} = \frac{2}{3} \Dot{M}\Omega R^{2}_{*}\eta_* \label{eq:j_dot_mesa}
\end{equation}
\end{ceqn}
which is a more convenient expression to be implemented in MESA because is based on values directly exposed during the simulations.

\section{Results} \label{sec_3}

\subsection{Mesh Models} \label{sec_mesh}
According to the formulation in the previous sections, the only two free parameters of our implementation are $B$ and $\Omega$. The numerical simulations traced the rotational history and $A(\isotope[7]{Li})$ of a $1\, \msun$ star for a variety of initial values for $B$ and $\Omega$ (see also Table \ref{tab:phy_mesa}). MESA uses the shellular approach \citep{Meynet1997} to account for hydrostatic effects of rotation into 1D stellar models.\par

\begin{table}
	\centering
	\caption{Summary of adopted physics in MESA \citep[based on][]{Choi2016}.}
	\label{tab:phy_mesa}
	\begin{tabular}{ll} 
		\hline
		Parameter & Adopted prescriptions and values\\
		\hline
		Solar Abundance & $X_{\odot}=0.7154, Y_{\odot}=0.2703, Z_{\odot}=0.0142$\\
		Equation of State & OPAL+SCVH+MacDonald+HELM+PC\\
		Opacity & OPAL Type I for log T $\geq$ 4 \\ & Ferguson for logT $\leq$ 4\\
		Reaction Rates & JINA REACLIB\\
		Boundary Conditions & ATLAS12; $\tau$=100 tables + photosphere\\
		Difussion & Track \isotope[1]{H}, \isotope[2]{He}, \isotope[7]{Li}, \isotope[7]{Be}\\
		Rotation & Differential rotation at PMS \& MS\\
		Convection & MLT + Ledoux, $\alpha_{{\rm MLT}}$ = 1.82\\
		Overshoot & time-dependent, diffusive, \\ & $f_{{\rm ov,core}}=0.0160$,\\ 
		& $f_{{\rm ov,sh}}=0.0174$\\
		Semiconvection & $\alpha_{{\rm sc}}=0.1$\\
		Thermohaline & $\alpha_{{\rm th}}=666$\\
		Rotational Mixing & Include SH, ES, GSF, SSI \& DSI\\
		Magnetic Effects & Magnetic braking based on idealized \\ & monopole field\\
		Magnetic Field & B(G) variable between [3.0 - 5.0]\\
		Mass Loss & activated, $\Dot{M}_{{\rm max}} = 10^{-3} \: \msun \: yr^{-1}$\\
		Angular Moment Loss & activated, $\Dot{J} = \frac{2}{3} \Dot{M}\Omega R^{2}_{A}$\\
		\hline
	\end{tabular}
\end{table}

We adopted solar-scaled abundances and assume \citet{Asplund2009} solar initial abundance ($Z = Z_{\odot, \mathrm{pr} = 0.0142}$). We also adopted the following nominal values to express stellar properties values in SI units, $\rsun = 6.957x10^{10}\, cm$ and $\msun = 1.988x10^{33}\, g$ which are consistent with IAU resolutions \citep{Mamajek2015}. For a detailed description of the physics adopted in this paper, we refer the reader to \citet{Choi2016}. That work was used as starting point for ours regarding the parameterization of the MESA project which was calibrated to reproduce the measured element abundances on the solar surface.\par

The models included rotation during the PMS as there are evidences that advocate for a strong established relationship between Li destruction and rotation on that phase (\citeauthor{Bouvier2016} \citeyear{Bouvier2016}, \citeyear{Bouvier2018}). We modeled rotation such as AML is computed as a result of the torque applied to the convection zone by a magnetically-coupled wind. Additionally, we did not take into account either the influence of internal magnetic fields or their existence during the T-Tauri phase. We adopted a simple and pragmatic theoretical approach to establish when the star is reaching the ZAMS phase. The selected criterion was based on the simultaneous existence of an extensive convective layer and a radiative core. From this moment on, the MB routine was activated, acting as an additional mechanism to those existing in the MESA evolutionary code that participated in the star AML. We assumed that the magnetic field did not vary its intensity throughout the evolution of the star until it reached the TAMS. In addition, other rotation effects were also considered during the evolution of the models: the AM transport from the radiative interior to the convective envelope and the AM redistribution associated with changes in the internal structure during the process of contraction to the MS.\par

In this study we focused on the indirect role played by the MB on the Li destruction. This is a consequence of its influence on the rotational history of solar-type stars. Our major objective was to check how MB could contribute to explain the evolution of Li in the Sun and other solar-type stars. A set of different scenarios with magnetic field strengths ranging between 3.0 and 5.0 G were exercised. We computed the evolution of $1\,\msun$ stellar models at solar initial metallicity with $\oomegac$ varying between $0.0084$ and $0.0336$. More advanced models for more complex arrangements of the magnetic field as well as the evolution of its intensity throughout the life of the star were left for later work.\par

As expressed by Eq.~\ref{eq:j_dot_mesa}, the amount of AML depends on $R_A$ (or alternatively on $\eta_*$), $\Omega$ and $\Dot{M}$. For large $\eta_*$ value, the star undergoes a significant deceleration. We controlled the value of $\eta_*$ in the models by varying the magnetic field ($B$) and the stellar rotation speed ($\Omega$). With regards to $\Dot{M}$ and as reported in Table \ref{tab:phy_mesa}, the empirical formula developed by Reimers \citep{Reimers1975} for stars in the asymptotic giant branch (AGB) was used for calculating the mass-loss. For a solar-type star the $\Dot{M}$ during MS is relatively small, about  $10^{-14}\msun \, yr^{-1}$ \citep{Noerdlinger2008}. \par

MESA assigns an $\Omega$ value for each cell $k$ ($\Omega_k$) which is adjusted so that the resulting angular momentum is retained after calculating the new mass of the cell $k$ ($m_k$) and its distance to the center of the star ($r_k$). After that, an AM value is assigned to each cell $k$ ($J_k$). At this point, our MB turns on, modifying $J_k$. This was done by providing an additional contribution ($\Dot{J}_{k}$). This contribution is the result of the external torque exerted by the magnetic field once it has been distributed among the different layers that make up the CZ as dictated by Eq.~\ref{eq:k_jdot}:\par
 
\begin{ceqn}
\begin{align}
    \Dot{J}_{k} &= \Dot{J}_*\;\frac{m^{}_{k} r^2_{k}}{m^{}_* r_*^2} \label{eq:k_jdot}
\end{align}
\end{ceqn}

\subsection{Li evolution without MB}

\begin{figure}
	\includegraphics[trim = 25mm 10mm 15mm 10mm, clip, width=\columnwidth]{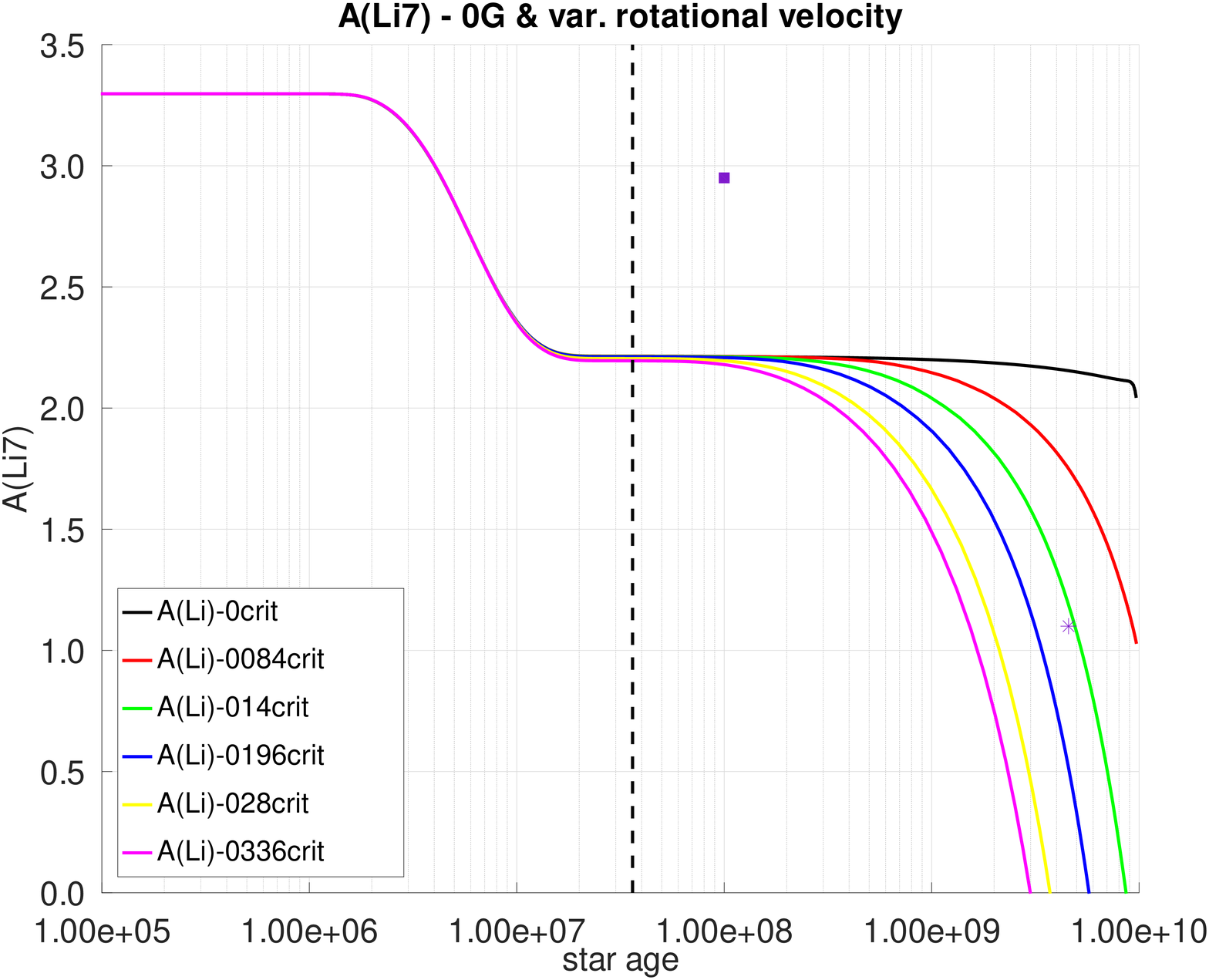}
    \caption{The evolution of surface \isotope[7]{Li} abundance relative to \isotope[1]{H}, as a function of time for several 1 $\msun$ models. The solid black line represents the reference model according to \citet{Choi2016}. The rest of lines are models which include PMS rotation with $\oomegac$ between 0.0084 and 0.0336, respectively. The purple star and square are surface Li abundances for the present-day Sun \citep{Asplund2009} and the average for the Pleiades cluster \citep{Sestito2005} respectively. The dashed vertical line makes reference to the ZAMS.}
    \label{fig:li_var_vel_0g}
\end{figure}

Figure \ref{fig:li_var_vel_0g} shows the temporal evolution of surface Li abundance for several $1\msun$ models initialized with different rotational velocities. The simulations took into account the effects of rotation and AML caused by stellar winds but not those of MB. The purple star and square are the surface Li abundances for the present-day Sun \citep{Asplund2009} and for the Pleiades, respectively \citep{Sestito2005}.\par

\begin{figure}
	\includegraphics[trim = 25mm 10mm 15mm 10mm clip,width=\columnwidth]{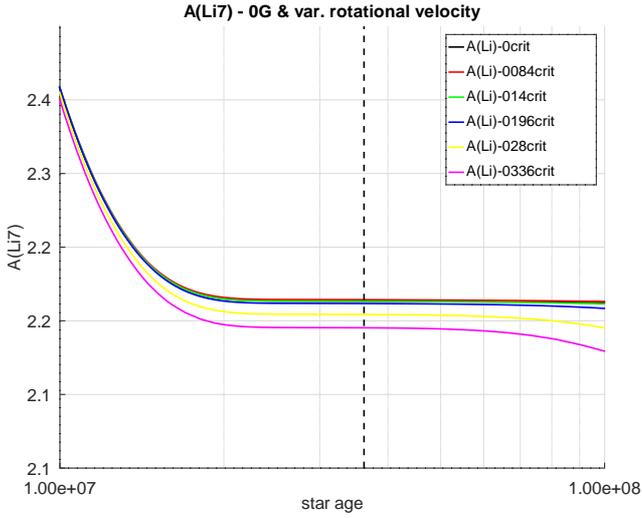}
    \caption {Similar to Figure \ref{fig:li_var_vel_0g} but zooming in on the ZAMS. The models with a higher initial rotational velocity reach already the ZAMS with a lower amount of Li measure on the star surface.}
    \label{fig:li_var_vel_0g_z1}
\end{figure}

\begin{figure}
	\includegraphics[trim = 10mm 10mm 15mm 10mm clip,width=\columnwidth]{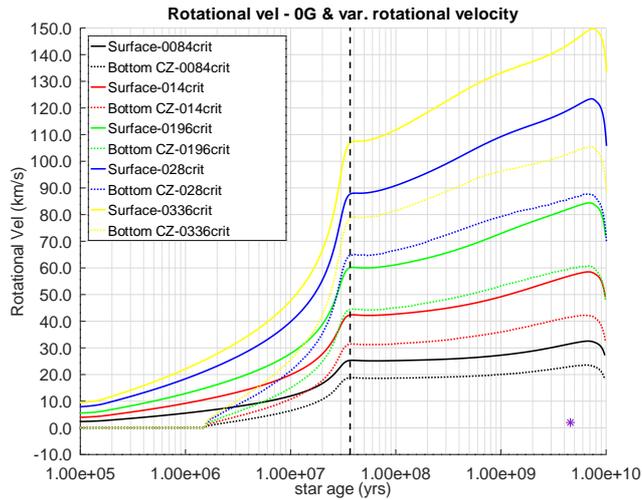}
    \caption{The evolution of angular velocity at surface (solid line) and at the bottom (dotted line) limit of the uppermost convective zone, as a function of time for several 1 $\msun$ models. The models includes PMS rotation with $\oomegac$ values between 0.0084 and 0.0336. The purple star is the surface angular velocity for the present-day Sun \citep{Gill2012}. The dashed vertical line makes reference to the ZAMS.}
    \label{fig:rot_vel_0g}
\end{figure}

Notice how the Li abundance in the star surface decreased over time for all simulated models. The solid black line represents the reference model that adopts the solar-calibrated envelope overshooting parameters as documented in \citet{Choi2016}. All models burned too much Li before the ZAMS and therefore did not match with the Pleiades average surface Li abundance. Also remarkable was the fact that there were barely any differences between the different models in terms of the abundance of Li for much of the PMS. Only after one million years was Li destroyed to an accentuated degree, since the necessary temperature was not reached at BCZ before. Afterwards the different models destroyed the Li in a very similar way due to two main reasons. On the one hand, the convective zones that the models developed had very similar sizes, so that the temperature at the BCZ was practically the same. On the other hand, the differential rotation between the core and the convective zone did not begin to develop until it reached $ \approx 10^6$ years, achieving its maximum difference in the ZAMS about $ \approx 10^7$ years. It was at this moment that the combined effects of turbulence and rotational difference should increase, leading to notable differences in the evolution of A(Li) (see Figure \ref{fig:li_var_vel_0g_z1}). Later, the reference model (black line) did not deplete Li efficiently on the MS and failed (again) to match the current solar surface Li abundance. The other models which included rotation during the PMS with values of $\oomegac$ between 0.0084 and 0.0336 were able to burn Li in a more realistic way. Among them only one (green line) was close to match the present-day Li abundance of the Sun but its rotational velocity was much higher (see Figure \ref{fig:rot_vel_0g}) than the $2\,\kms$ of the Sun \citep{Gill2012}. \par

For much of the PMS the star rotated as a solid body (see Figure \ref{fig:rot_vel_0g}) and this was because the star had a full convective interior. It was not until the end of the Hayashi track that the star began to develop a radiative core. It was in this stage when a difference in angular velocity appeared between the upper and lower limits of the radiative and convective zones respectively. The degree of differential rotation was directly influenced by the initial $\Omega$. As the models were initialized with a higher angular velocity, the difference in velocity between the BCZ and the star surface was accentuated; the higher the initial velocity, the bigger the velocity gradient between the bottom and top limits of the CZ. As a consequence, the turbulence strength located at the BCZ increased so that Li could reach regions with temperatures about $\tli$ where it was finally burned and destroyed (see Figure  \ref{fig:li_var_vel_0g}). Other investigations \citep{Bouvier2018, Baraffe2017}  point to a tendency diametrically opposed to the one exposed here, i.e. the faster the rotation speed, the greater the abundance of Li on the surface of the star. \par

\begin{figure}
	\includegraphics[trim = 10mm 10mm 15mm 10mm, clip,width=\columnwidth]{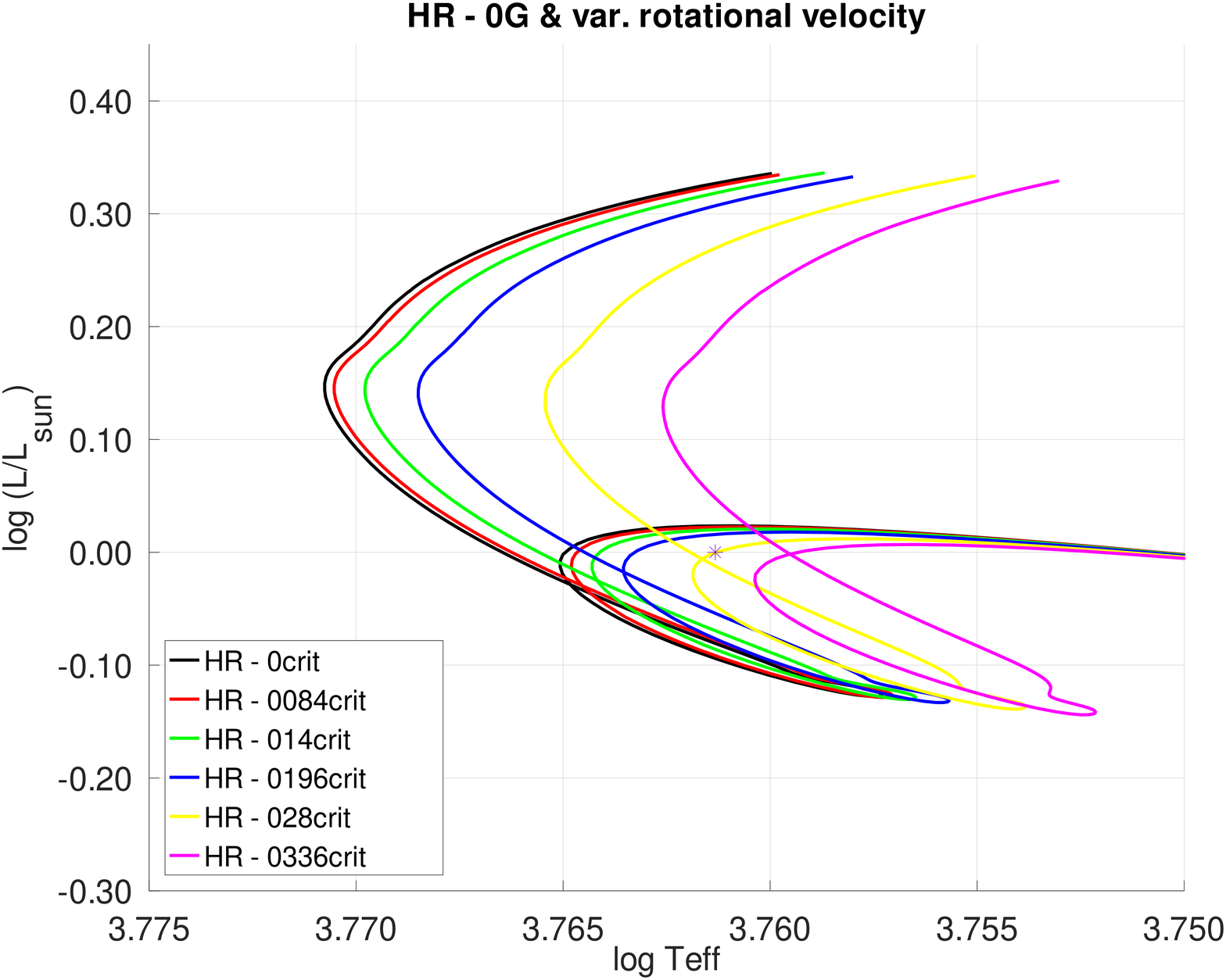}
    \caption{An example solar 1$\msun$ grid of stellar evolutionary track from PMS till TAMS covering a wide range of angular velocities. The rotation is activated in the models in the PMS and those models reach before the ZAMS and at a lower $\teff$ than the non-rotating one (solid black line). The luminosity is expressed in terms of $\lsun$.}
    \label{fig:hr_var_vel_0g}
\end{figure}

Other well-known structural effects of rotation are the decrease of the effective temperature ($\teff$) and to a less extent of stellar luminosity ($L$). Both effects can be observed graphically in the HR diagram of Figure \ref{fig:hr_var_vel_0g} which shows a zoomed-in view of evolutionary tracks from the ZAMS until the TAMS for several $1\msun$ models initialized with different rotational velocities. If we compare the non-rotating model (black solid line) with the rotating ones we can recognize that at the end of the PMS, the latter reach the ZAMS with a lower $\teff$ than the former. These results are in line with those of previous studies \citep[see e.g. ][]{Eggenberger2012,Piau2001,Pinsonneault1989}.\par

\subsection{Li evolution with MB}
\begin{figure}
	\includegraphics[trim = 25mm 10mm 15mm 10mm, clip,width=\columnwidth]{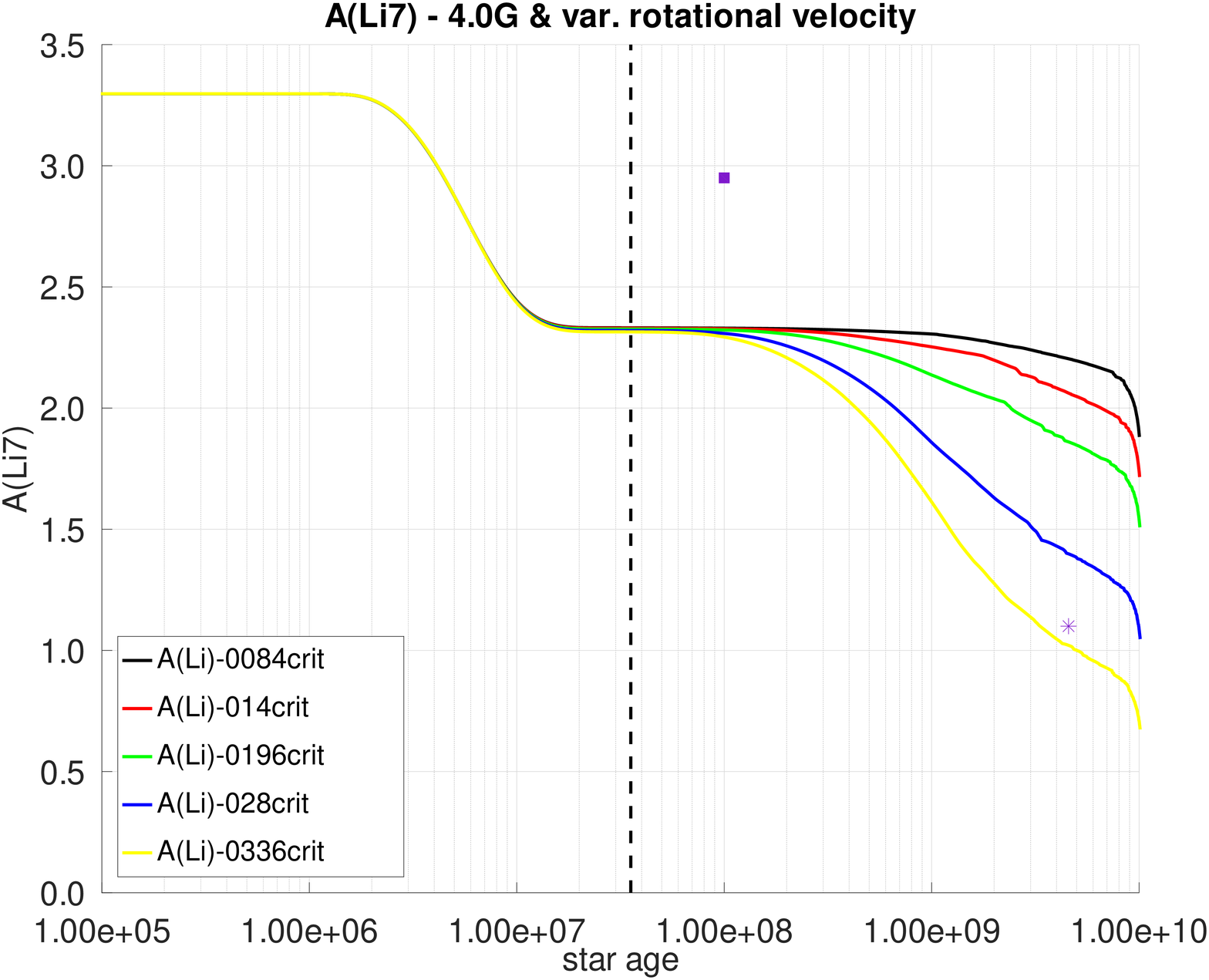}
    \caption{The evolution of surface \isotope[7]{Li} abundance relative to \isotope[1]{H}, as a function of time for several 1 $\msun$ models. The models include a magnetic field with an intensity of 4G and PMS rotation with $\oomegac$ between 0.0084 and 0.0336, respectively. The purple star and square are surface Li abundance for the present-day Sun \citep{Asplund2009} and the Pleiades cluster \citep{Sestito2005} respectively. The dashed vertical line makes reference to the ZAMS.}
    \label{fig:li_var_vel_4_0g}
\end{figure}

Figure \ref{fig:li_var_vel_4_0g} shows the temporal evolution of surface Li abundance for several 1 $\msun$ models. Those models were initialized with different rotational velocities and took into consideration the effects of MB caused by a magnetic field of intensity 4G. If we compare it with Figure \ref{fig:li_var_vel_0g} in which the effects of MB were neglected, we notice how the profiles of Li abundance were altered during PMS and MS. During the PMS we can describe the effect as modest, somewhat expected and in line with the fact that the AML caused by MB (see Eq.~\ref{eq:j_dot}) depends directly on the mass loss rate. If we take into account that for solar-type stars the models predict a modest total mass loss rate, that value is even much lower in this phase. On the contrary, during the MS it is observed that the AML is much more significant, causing a smaller amount of Li to be destroyed.\par

\begin{figure}
	\includegraphics[trim = 10mm 10mm 15mm 10mm, clip,width=\columnwidth]{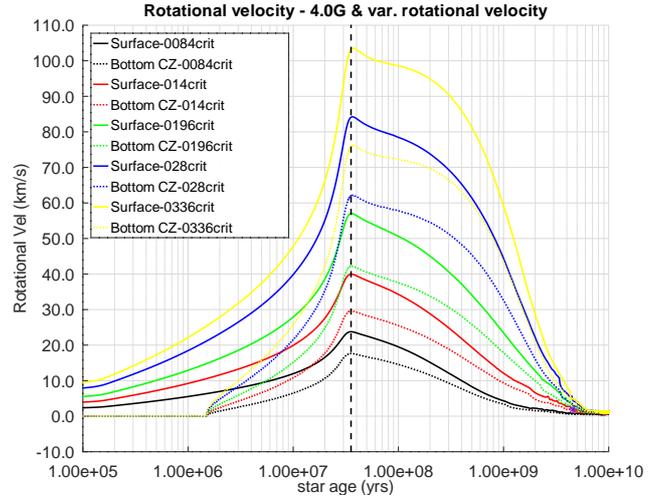}
    \caption{The evolution of surface rotational velocity, as a function of time for several 1 $\msun$ models. The models include a magnetic field with an intensity of 4\,G, PMS rotation with $\oomegac$ between 0.0084 and 0.0336, respectively and MB. The purple star is the surface angular velocity for the present-day Sun \citep{Gill2012}. The dashed vertical line makes reference to the ZAMS.}
    \label{fig:rot_vel_4g}
\end{figure}

\begin{figure}
	\includegraphics[trim = 10mm 10mm 15mm 10mm, clip,width=\columnwidth]{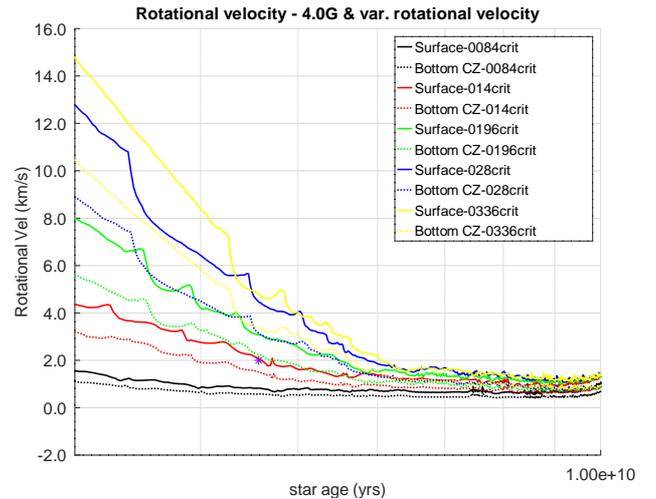}
    \caption{Similar to Figure \ref{fig:rot_vel_4g} but now showing in detail the surface rotational velocity as the star approaches TAMS.}
    \label{fig:rot_vel_4g_z1}
\end{figure}

The effect of the MB routine can be seen even more clearly in Figures~\ref{fig:rot_vel_4g}, \ref{fig:rot_vel_4g_z1} \& Appendixes \footnote{The appendices comprise a series of grids as a function of time and for several 1 $\msun$ models in which on the one hand, the evolution of surface \isotope[7]{Li} abundance relative to \isotope[1]{H} for both variable magnetic field intensities and angular velocities and on the other hand, the evolution of surface rotational velocity are shown.}. In those figures we represented the rotation profiles for the surface of the stars and for the bottom of the convective envelope. They are 1 $\msun$ models initialized with different rotational velocities and considering the influence of MB. Similarly to the $A(\isotope[7]{Li})$ evolution profiles commented in the previous paragraph, the effect of the routine were visible once the ZAMS was reached. If we compare the evolution of the curves presented here with those of Figure \ref{fig:rot_vel_0g} we see how the star, instead of keep increasing $\Omega$, began to slow down after having reached its maximum in the passage through the ZAMS. Notice that at this point the angular velocities at the surface of the star and at the BCZ reached their maximum difference. On the other hand, the MB effect caused the angular velocities between both zones of the star to decrease until, for an age close to that of the Sun (see Figure \ref{fig:rot_vel_4g_z1}), the star practically rotated like a rigid solid. These results were also consistent with those obtained by \citet{Eggenberger2010} as far as the effect of the magnetic field, in particular its influence on the loss of angular momentum, has on the rotational velocity of the star.\par

In a similar way we also observed that the models with lower angular velocity generally ended up exhibiting higher values for the Li abundance on the surface (see Figures~ \ref{fig:li_var_vel_4_0g}, \ref{fig:grid_li_var_vel} \& \ref{fig:grid_li_var_g}). In none of those cases we did obtain values of Li on the surface higher than those shown by the model without rotation.\par

\begin{figure}
	\includegraphics[trim = 10mm 10mm 15mm 10mm, clip,width=\columnwidth]{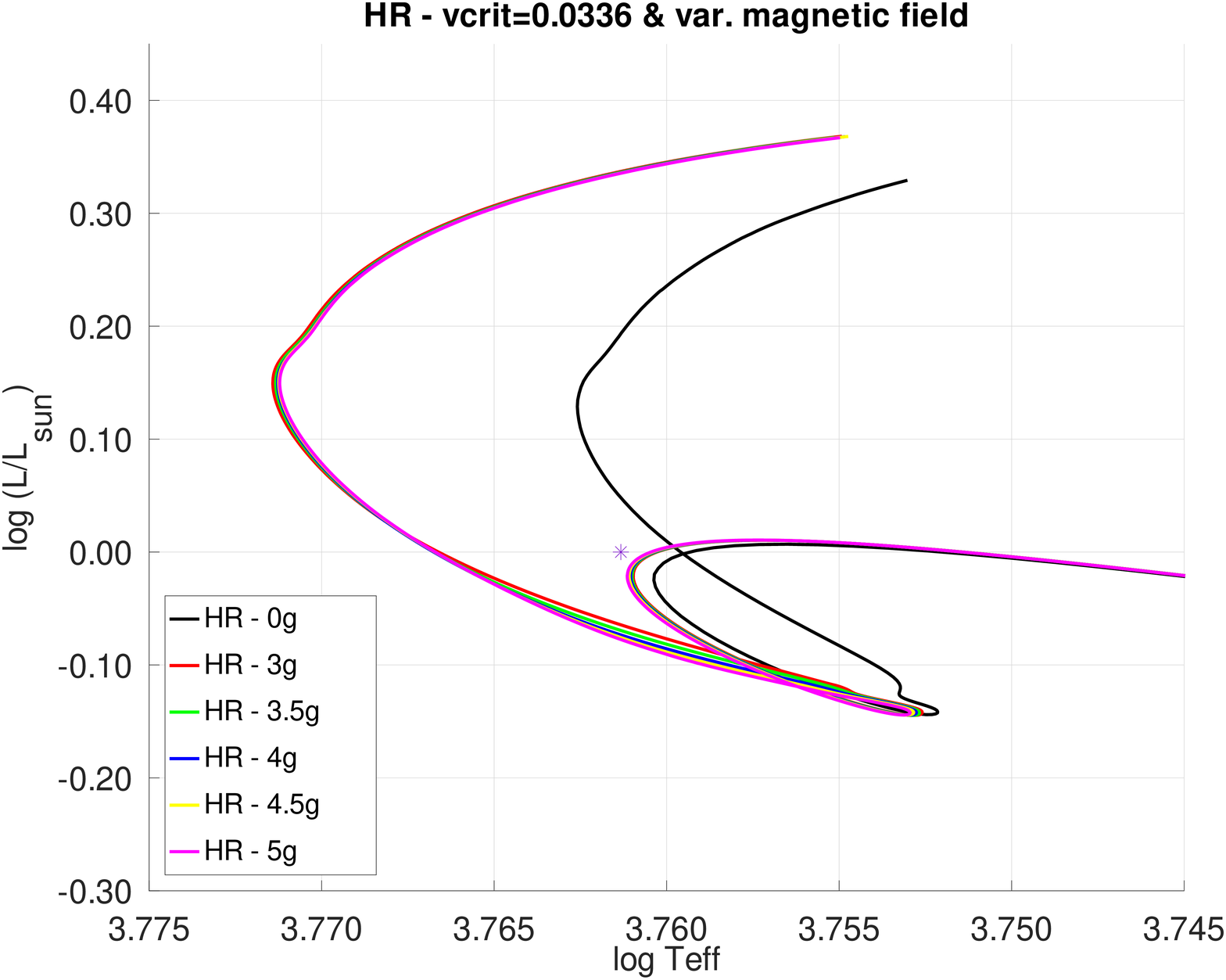}
    \caption{Similar to Figure \ref{fig:hr_var_vel_0g} but now showing in detail the effects of magnetic braking on the evolutionary tracks for different magnetic field strengths and $\oomegac=0.0336$. The presence of a magnetic field produces hotter stars due to the influence of the magnetic braking on the rotational velocity of the star.}
    \label{hr_vc_0336_var_g_z1}
\end{figure}

The MB also left its mark on the HR diagram by significantly affecting the $\teff$ of the star. To visualize this effect we take as reference Figure \ref{hr_vc_0336_var_g_z1} in which all models were started with the same value $\oomegac=0.0336$ but the intensity of the simulated magnetic field was different. Notice that the models with MB produced hotter stars due to its influence. The lower speed with which the star rotated due to the MB effect caused the increase of the $\teff$, being this difference of practically $95\,\Kelvin$ between the simulated models with $0.0\,\Gauss$ and $5.0\,\Gauss$ respectively for $log(\llsun)=0$.

\begin{figure}
	\includegraphics[trim = 15mm 10mm 15mm 10mm, clip,width=\columnwidth]{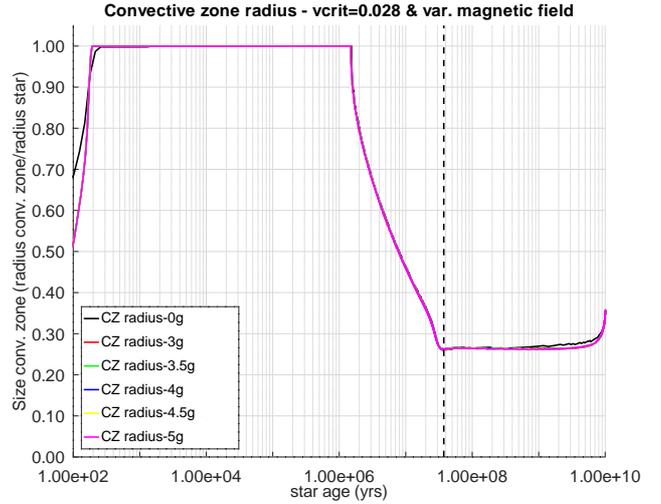}
    \caption{The evolution of convective zone size as a function of time for several 1 $\msun$ models. The models were all initialized with $\oomegac=0.0336$ and magnetic field strengths vary between $0.0\,\Gauss$ and $5.0\,\Gauss$. The dashed vertical line makes reference to the ZAMS.}
    \label{fig:cz_vc_028_var_b}
\end{figure}

\begin{figure}
	\includegraphics[trim = 15mm 10mm 15mm 10mm, clip,width=\columnwidth]{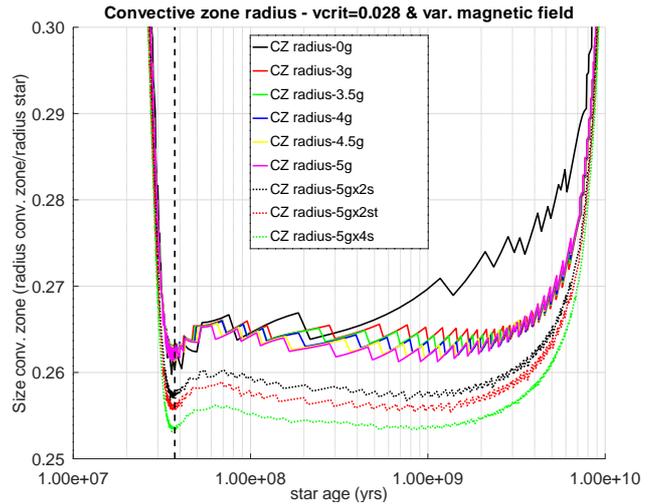}
    \caption{Similar to Figure \ref{fig:cz_vc_028_var_b} but now showing in detail the size of the convection zone in the models from ZAMS up to TAMS. As the intensity of the magnetic field increases, the size of the convective zone decreases. The first three pointed curves (below 0.26), have been artificially shifted downwards to illustrate the impact of the temporal (2x in time step and number of shells, red-dotted line) and spatial (2x and 4x number of shells, black and green-dotted lines, respectively) resolution on the numerical solution for the 5G case.}
    \label{fig:cz_vc_028_var_b_z1_new}
\end{figure}

As described in section \ref{sec_mesh}, the MB routine distributed the total amount of AML calculated according to Eq.~\ref{eq:k_jdot} among the different layers that composed the CZ. In Figure \ref{fig:cz_vc_028_var_b} we can observe the evolution of the most external CZ normalized with respect to the radius of the star for several 1 $\msun$ models. The models were all initialized with $\oomegac=0.0336$ and magnetic field strengths varying between $0.0\,\Gauss$ and $5.0\,\Gauss$. In accordance with the established models of stellar evolution, in a solar-type star the CZ covers practically all of it for a large part of the PMS. As it approaches the ZAMS, the CZ is decreasing as a consequence of the appearance of a radiative core and maintains an approximately constant radius until the final stage of the MS. In this point it increases significantly as a response to the generalized expansion of the star's radius. Regarding the effect of MB on the size of the CZ, we observe that as the intensity of the magnetic field increased, the size of the CZ decreased (see Figure \ref{fig:cz_vc_028_var_b_z1_new}). The radiative core pushed outward to include a rapidly increasing fraction of the stellar mass, making that the temperature at the CZ base dropped below $\tli$. This effect was most evident during MS. The decrease in the size of the CZ was in line with the fact that less Li was destroyed by causing less stellar material to reach areas with temperatures above $\tli$.\par

Notice the jagged appearance in Figure~\ref{fig:cz_vc_028_var_b_z1_new}. We suspected a numerical origin for the wriggles. In order to check it out, we computed models for all the physical options considered with higher temporal (time step) and spatial (number of shells) resolutions. For the sake of visual clarity, we illustrated the impact on the numerical solutions only for the 5G case, although the effects are the same in all cases. These are indicated as dotted lines in the figure, shifted-down just to become visible. Note that the larger the spatial resolution the smoother is the curve. The effect is less significant for the temporal resolution. The curve with 2x in both spatial and temporal resolution shows no difference with the case of 2x in only the number of shells. The 4x in spatial resolution points to the most appropriate number of shells to fine-track the evolution of the convective zone. However such high resolution implies to significantly increase the computation time, with no effect in the final conclusions of this work. Indeed, the size of the zigzag features is around $10^{-3}$,  which is far below the convection timescale and Li abundances measured in the stellar surface. It is thus save to neglect such effects by using the original time/space resolution.

\par

\begin{figure}
	\includegraphics[trim = 5mm 10mm 20mm 10mm, clip,width=\columnwidth]{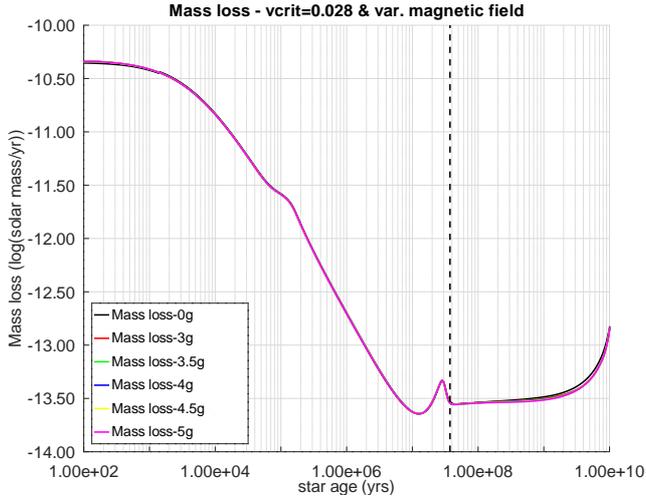}
    \caption{The evolution of mass loss $\Dot{M}$ as a function of time for several 1 $\msun$ models. The models include different magnetic field intensity between $0.0\,\Gauss$ and $5.0\,\Gauss$ and PMS rotation with $\oomegac=0.028$. The dashed vertical line makes reference to the ZAMS.}
    \label{fig:mdot_vc_028_var_b}
\end{figure}

\begin{figure}
	\includegraphics[trim = 5mm 10mm 20mm 10mm, clip,width=\columnwidth]{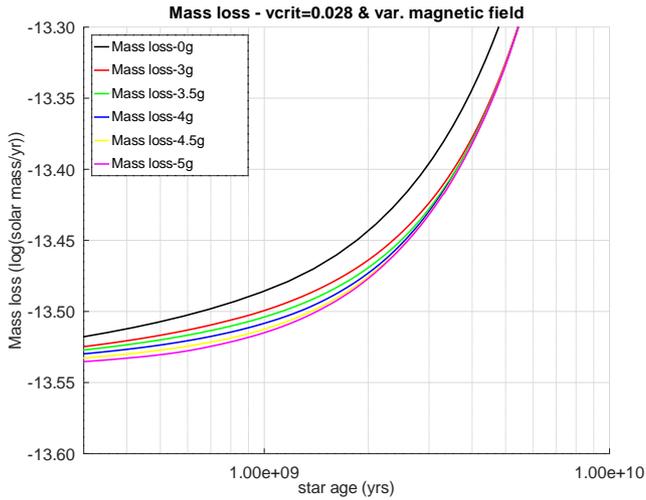}
    \caption{The evolution of mass loss $\Dot{M}$ as the star is approaching TAMS, as a function of time for several 1 $\msun$ models. The models include a variable magnetic field intensity between $0.0\,\Gauss$ and $5.0\,\Gauss$ and PMS rotation with $\oomegac=0.028$. The stronger the magnetic field, the lower $\Dot{M}$.}
    \label{fig:mdot_vc_028_var_b_z1}
\end{figure}

\begin{figure}
	\includegraphics[trim = 15mm 10mm 15mm 10mm, clip,width=\columnwidth]{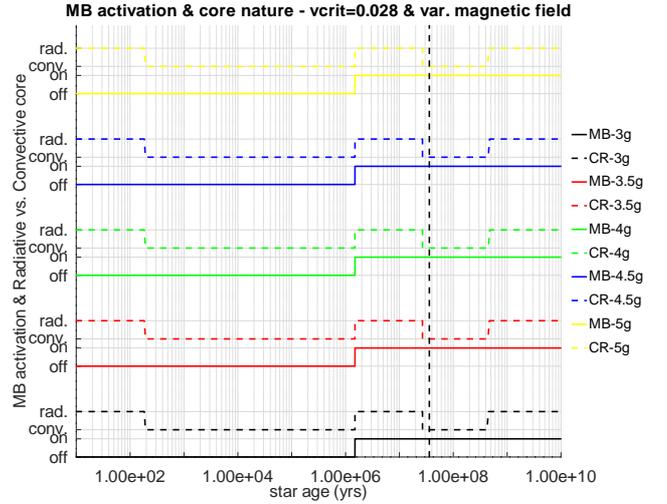}
    \caption{The activation of the magnetic braking routine as a function of the presence of a radiative core. The models include a magnetic field with an intensity ranging between $3.0\,\Gauss$ and $5.0\,\Gauss$ and PMS rotation with $\oomegac=0.0228$. The solid lines signal the magnetic braking routine activation (on) and deactivation (off). The horizontal dashed lines inform about the star's core nature: radiative (rad) or convective (conv). By implementation decision, once the routine is activated, it remains on even if the star's core nature change to convective. The dashed vertical line makes reference to the ZAMS.}
    \label{fig:mb_act_var_vel_vc_028}
\end{figure}

In Figure \ref{fig:mdot_vc_028_var_b} we can see the evolution of $\Dot{M}$ during PMS and MS for several 1 $\msun$ models. The models were all initialized with $\oomegac=0.0336$ and magnetic field strengths varying between $0.0\,\Gauss$ and $5.0\,\Gauss$. In the first stages of PMS was where the greatest mass loss was concentrated and this diminished as it approached the ZAMS. Because the star also decreased its radius during the PMS, it increased $\Omega$ obeying the principle of conservation of AM. When reaching the ZAMS, the stellar radius remained more or less stable for much of the MS (except for its final stage, see Figure \ref{fig:mdot_vc_028_var_b_z1}) but continued to lose mass whilst increasing $\Omega$, although in a less aggressive way if we compare it with the PMS. However, as a consequence of both the appearance of the radiative core during Henyey track and the existence of a CZ (see Figure \ref{fig:mb_act_var_vel_vc_028}), the MB routine was activated, causing the angular velocity of the star to begin to decrease along the entire MS. The more intense the magnetic field, the greater the braking effect.\par 

\subsection{Alternative Li evolution with MB}
\begin{figure}
	\includegraphics[trim = 25mm 10mm 15mm 10mm, clip,width=\columnwidth]{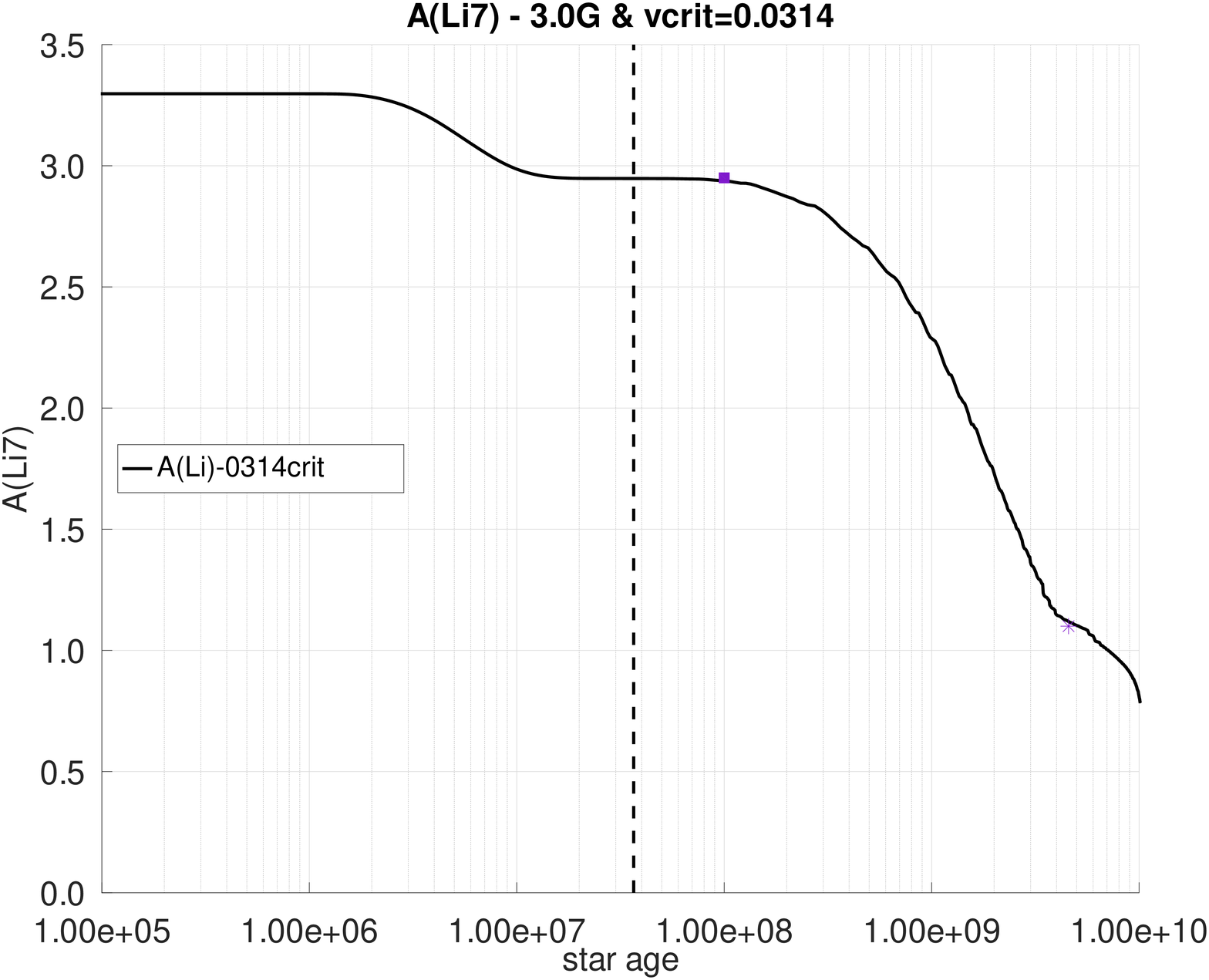}
    \caption{The evolution of surface \isotope[7]{Li} abundance relative to \isotope[1]{H}, as a function of time for a 1 $\msun$ model. The model include a magnetic field with an intensity of 3\,G and PMS rotation with $\oomegac= 0.0314$, respectively. MLT and overshooting parameters have been set to $\alpha_\mathrm{MLT}=1.70$, $f_\mathrm{ov,core}=0.016$, $f_\mathrm{ov,sh}=0.002$. The purple star and square are surface Li abundance for the present-day Sun \citep{Asplund2009} and the Pleiades cluster \citep{Sestito2005} respectively. The dashed vertical line makes reference to the ZAMS.}
    \label{fig:li_3_0g_0314vc}
\end{figure}

Up to this point, the simulations of the different models were based on the parameterization gathered in Table \ref{tab:phy_mesa}, which in turn was adopted from \citet{Choi2016}. If we recall the evolution of the Li shown in Figures~\ref{fig:li_var_vel_0g} and \ref{fig:li_var_vel_4_0g}, we highlighted that on both cases too much Li was burned before reaching the ZAMS and therefore did not match with the Pleiades average surface Li abundance. On the other hand, using a parameterization slightly different from the one used until now in which the parameters of convection and overshooting have been readjusted according to Table \ref{tab:phy_alt_mesa}, the simulations could reproduce more faithfully both the Pleiades cluster and the Sun Li abundances (see Figure \ref{fig:li_3_0g_0314vc}).\par

\begin{table}
	\centering
	\caption{Alternative MTL and overshooting parameters.}
	\label{tab:phy_alt_mesa}
	\begin{tabular}{ll} 
		\hline
		Parameter & Adopted prescriptions and values\\
		\hline
		Convection & MLT + Ledoux, $\alpha_\mathrm{MLT}$ = 1.70\\
		Overshoot & time-dependent, diffusive, $f_\mathrm{ov,core}=0.016$, \\ & $f_\mathrm{ov,sh}=0.002$\\
		\hline
	\end{tabular}
\end{table}

\begin{figure}
	\includegraphics[trim = 10mm 10mm 15mm 10mm, clip,width=\columnwidth]{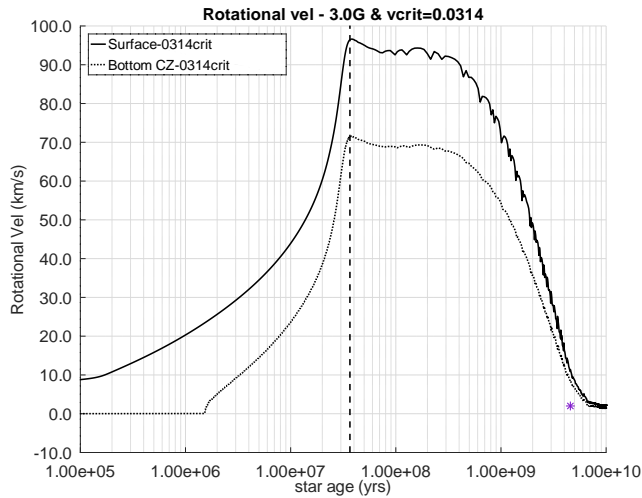}
    \caption{The evolution of surface rotational velocity, as a function of time for a 1 $\msun$ model. The model include a magnetic field with an intensity of 3\,G, PMS rotation with $\oomegac=0.0314$, parameters in Table \ref{tab:phy_alt_mesa} and MB. The purple star is the surface angular velocity for the present-day Sun \citep{Gill2012}. The dashed vertical line makes reference to the ZAMS.}
    \label{fig:rot_vel_var_vel_mlt_3_0g}
\end{figure}

Similarly, the evolution of the angular velocity for this new configuration is shown in Figure \ref{fig:rot_vel_var_vel_mlt_3_0g}. Here we did not yet reproduce correctly the angular velocity of the Sun. It became evident that the influence of the free, relatively arbitrary, parameters associated with MLT significantly conditioned the evolution of the Li abundance. However, without the inclusion of MB it was not possible to fit it for the Pleiades and the Sun with the same evolutionary track. This fact opens the door for other solar-compatible parameterizations to reproduce the observations in young stellar clusters and for the Sun. \par

\section{Conclusions and future works} \label{sec_conclusions}
We have shown, through the different simulated stellar models, that the effects induced by the combination of both rotation and magnetic braking mechanisms offer a plausible way to reconcile the observational data with the theoretical models. The latter is of major importance both for the transport of AM and for the transport of chemical elements. We are also well aware that we are still far from understanding the exact physical mechanisms that govern these processes, so it is necessary to continue to delve into these areas of study. Particularly the study of the evolution of the magnetic fields during the PMS and MS and their impact on AML. \par

Future and improved implementations of our routine will be made to study the results on tracks, surface Li composition, rotation, etc. It is likely that in a rotating stellar model simulated with MB from the beginning, the differential rotation is very much reduced and therefore Li abundances observed in young stars could be properly explained. We propose that this result could be achieved by a law of interdependence between $\Omega$ and $B$. So that, during the PMS, when the star rotates faster, the MB effect is more efficient. On the other hand, during the MS, when the start slows down, the MB will become less intensive. This approach would establish a self-regulating mechanism over the angular velocity of the star that would end up directly influencing the Li evolution. It is equally important to understand better the general role of mass loss in AML and mostly in cooler, low-mass stars. Nowadays it is challenging to determine the terminal velocity of the stellar wind for this type of stars which plays a key role on the AML.\par

We emphasize as well that our models failed to match at the same time the observed solar Li abundance and $\Omega$. We could not, therefore, ascertain to have correctly modeled the rotational history of the Sun. In view of these shortcomings in our models, we must analyse the results obtained with caution and not draw any premature conclusions. We have also shown how an alternative MLT parametrizations could produce results in line with the observations. The following conclusions have been supported in the course of this work:
\begin{enumerate}
    \item Inclusion of the magnetic field leads to cooler models and lower Li depletion in the MS.
    \item A combination of rotation during the PMS and MB effect during MS produces different, potentially more promising behavior than those produced by standard models. Thus, our approach points to reproduce the observed A(Li) and the solar rotational velocity at the same time.
    \item $\teff$ from standard evolutionary tracks represent upper limits since these models do not take into consideration the magnetic braking effect nor rotation.
    \item The convective zone extension decreases when the intensity of the magnetic field increases.
    \item MB during PMS and/or adjustment of MLT overshooting and $\amlt$ free parameters seems to be also required for explaining Li abundances in young clusters.
\end{enumerate}

The next steps will be focused on understanding how magnetic fields are linked to rotation, how their topologies are and how they evolve in time in order to go beyond the treatment on the MB routine that has been done in this work.\par

\section*{Acknowledgements}
We are pleased to acknowledge Jieun Choi her kindness in responding to the different questions about her work which were aimed to reproduce the results obtained as far as Li is concerned, and also for allowing us access to the MESA files she used. Also we very much appreciate the expert support provided by Elisa Delgado-Mena during the process review of this document. Similarly, we wish also to thank Matteo Cantiello and Bill Paxton for their very valuable support in the development of the magnetic braking routine in MESA. Finally, we wish to acknowledge the generous work and help offered by the MESA community. Authors acknowledges funding support from Spanish public funds (including FEDER fonds) for research under projects ESP2017-87676-C5-2-R and ESP2017-87676-C5-5-R. JCS also acknowledges support from project RYC-2012-09913 under the 'Ram\'on y Cajal' program of the Spanish Ministry of Science and Education.

\section*{Software}
The simulations used in this paper been executed with the MESA release 10398 \footnote{\url{http://mesa.sourceforge.net/release/2018/03/21/r10398.html}}. The different figures have been generated making use of GNU Octave 5.1.0. \footnote{\url{https://www.gnu.org/software/octave/}} 




\bibliographystyle{mnras}
\bibliography{mblithium}




\appendix

\section{Mesh models visualization}
The following figures comprise a series of grids as a function of time and for several 1 $\msun$ models in which on the one hand, the evolution of surface \isotope[7]{Li} abundance relative to \isotope[1]{H} for both variable magnetic field intensities and angular velocities and on the other hand, the evolution of surface rotational velocity are shown.

\begin{figure*}
    \centering
    \begin{subfigure}[h]{0.47\textwidth}
    \includegraphics[trim = 25mm 10mm 15mm 10mm, clip,width=\textwidth]{figures/li_var_vel_0_0g.eps}
    \label{fig:subim1}
    \end{subfigure}
    \begin{subfigure}[h]{0.47\textwidth}
    \includegraphics[trim = 25mm 10mm 15mm 10mm, clip,width=\textwidth]{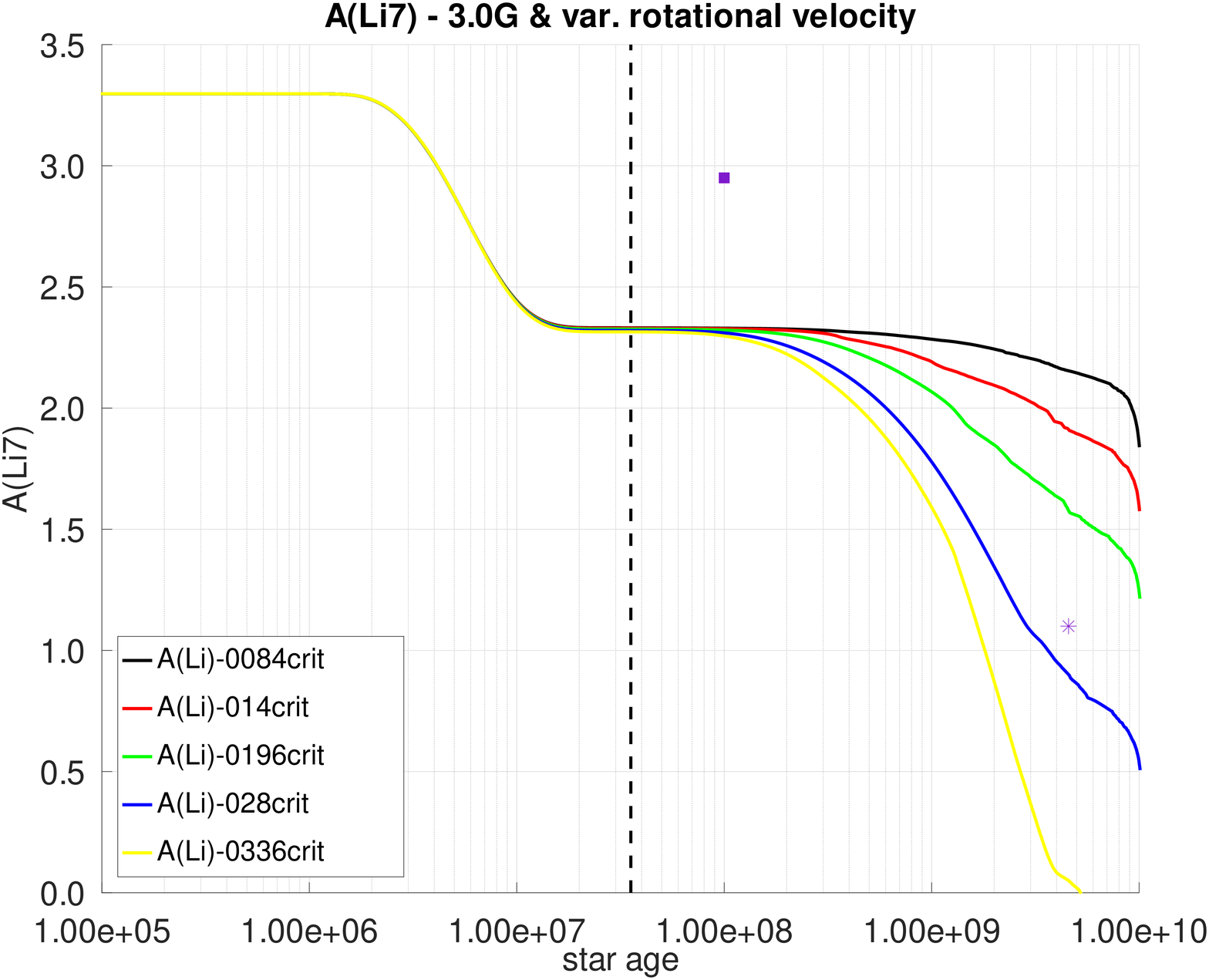}
    \label{fig:subim2}
    \end{subfigure}
    \begin{subfigure}[h]{0.47\textwidth}
    \includegraphics[trim = 25mm 10mm 15mm 10mm, clip,width=\textwidth]{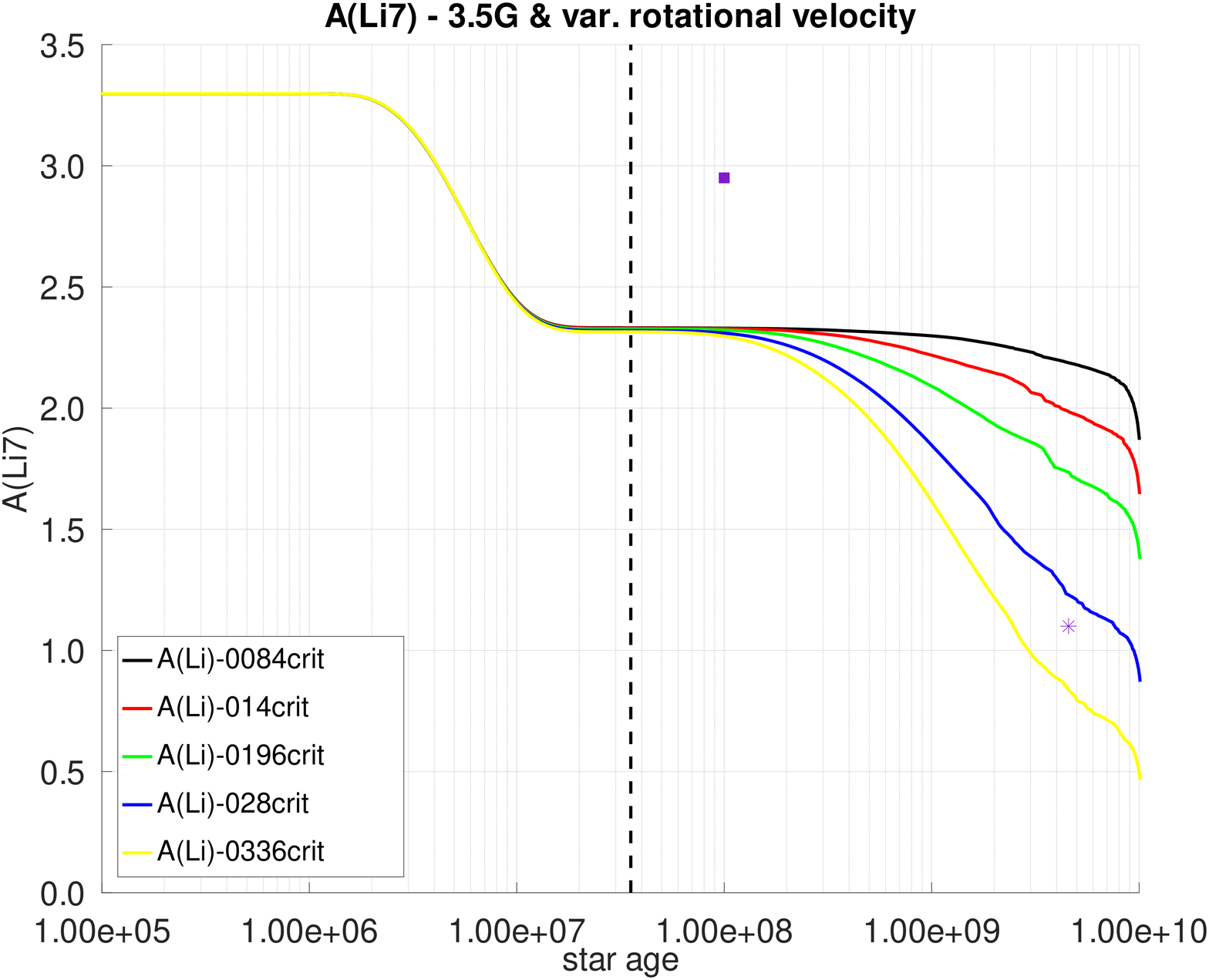}
    \label{fig:subim3}
    \end{subfigure}
    \begin{subfigure}[h]{0.47\textwidth}
    \includegraphics[trim = 25mm 10mm 15mm 10mm, clip,width=\textwidth]{figures/li_var_vel_4_0g.eps}
    \label{fig:subim4}
    \end{subfigure}
    \begin{subfigure}[h]{0.47\textwidth}
    \includegraphics[trim = 25mm 10mm 15mm 10mm, clip,width=\textwidth]{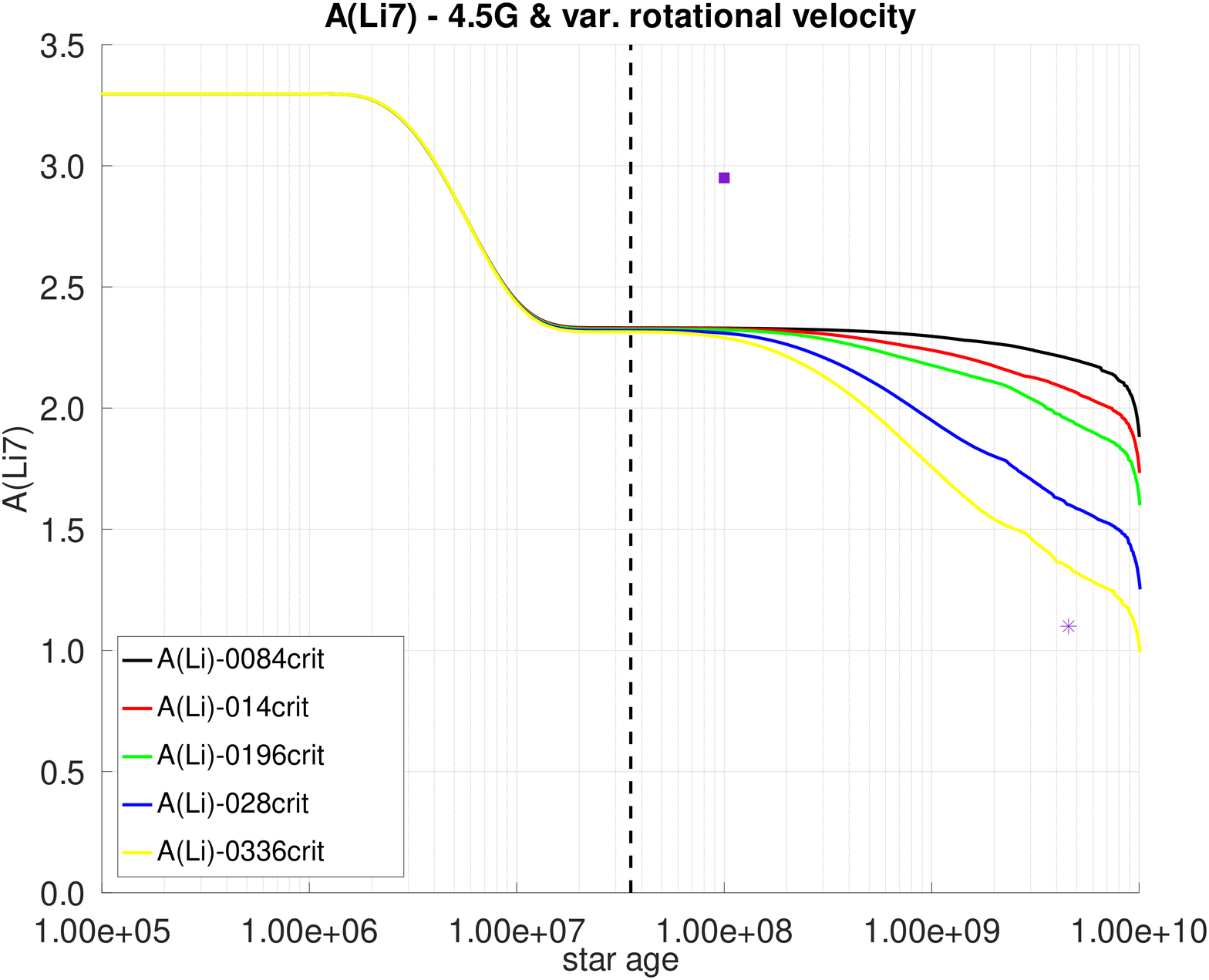}
    \label{fig:subim5}
    \end{subfigure}
    \begin{subfigure}[h]{0.47\textwidth}
    \includegraphics[trim = 25mm 10mm 15mm 10mm, clip,width=\textwidth]{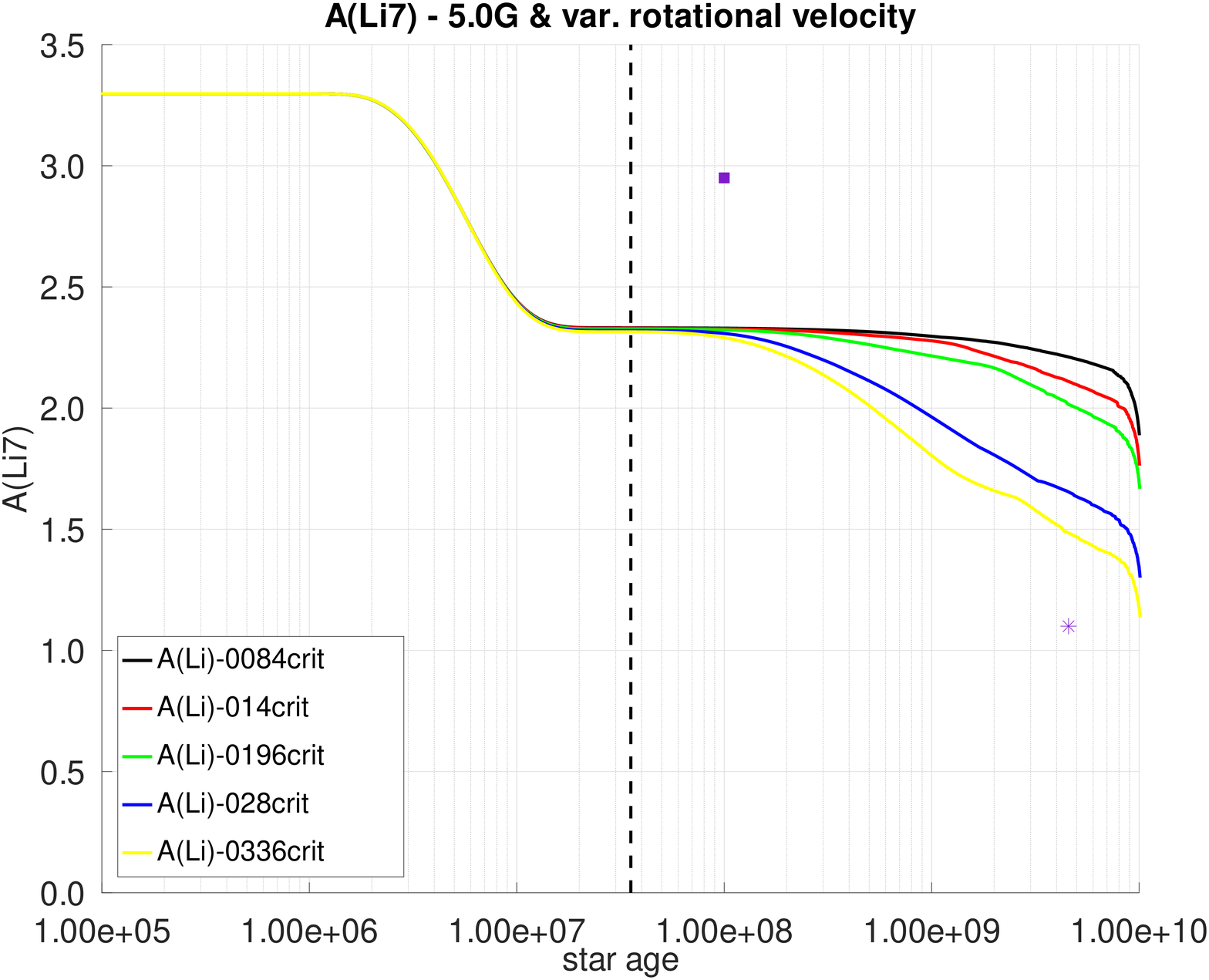}
    \label{fig:subim6}
    \end{subfigure}
\caption{Grid showing the evolution of surface \isotope[7]{Li} abundance relative to \isotope[1]{H}, as a function of time for several 1 $\msun$ models. Each figure shows a set of models in which the magnetic field with intensity has been fixed and $\oomegac$ varies between 0.0084 and 0.0336, respectively. The purple star and square are surface Li abundance for the present-day Sun \citep{Asplund2009} and the Pleiades cluster \citep{Sestito2005} respectively. The dashed vertical line makes reference to the ZAMS.}
\label{fig:grid_li_var_vel}
\end{figure*}
\par

\begin{figure*}
    \centering
    \begin{subfigure}[h]{0.47\textwidth}
    \includegraphics[trim = 25mm 10mm 15mm 10mm, clip,width=\textwidth]{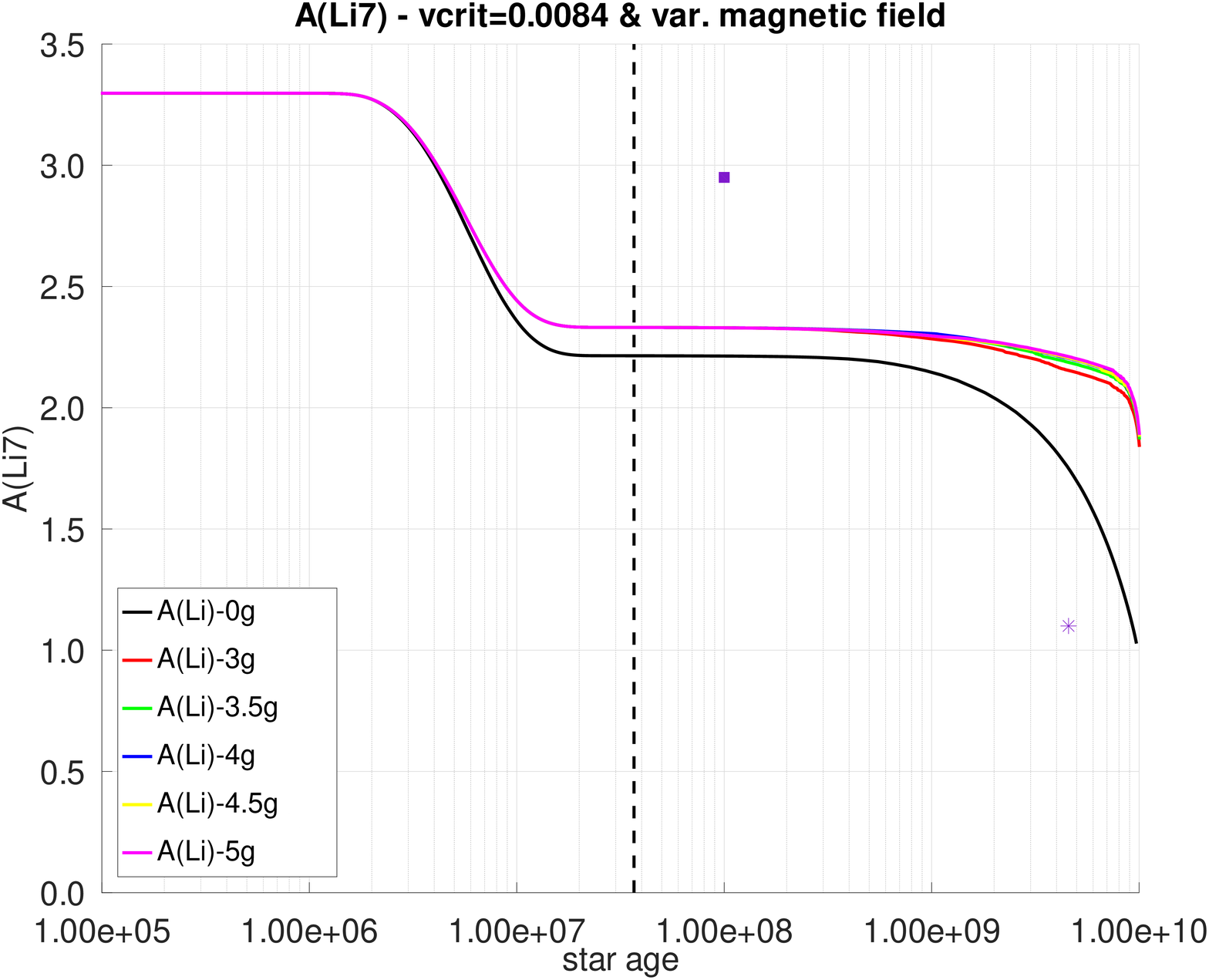}
    \label{fig:subim21}
    \end{subfigure}
    \begin{subfigure}[h]{0.47\textwidth}
    \includegraphics[trim = 25mm 10mm 15mm 10mm, clip,width=\textwidth]{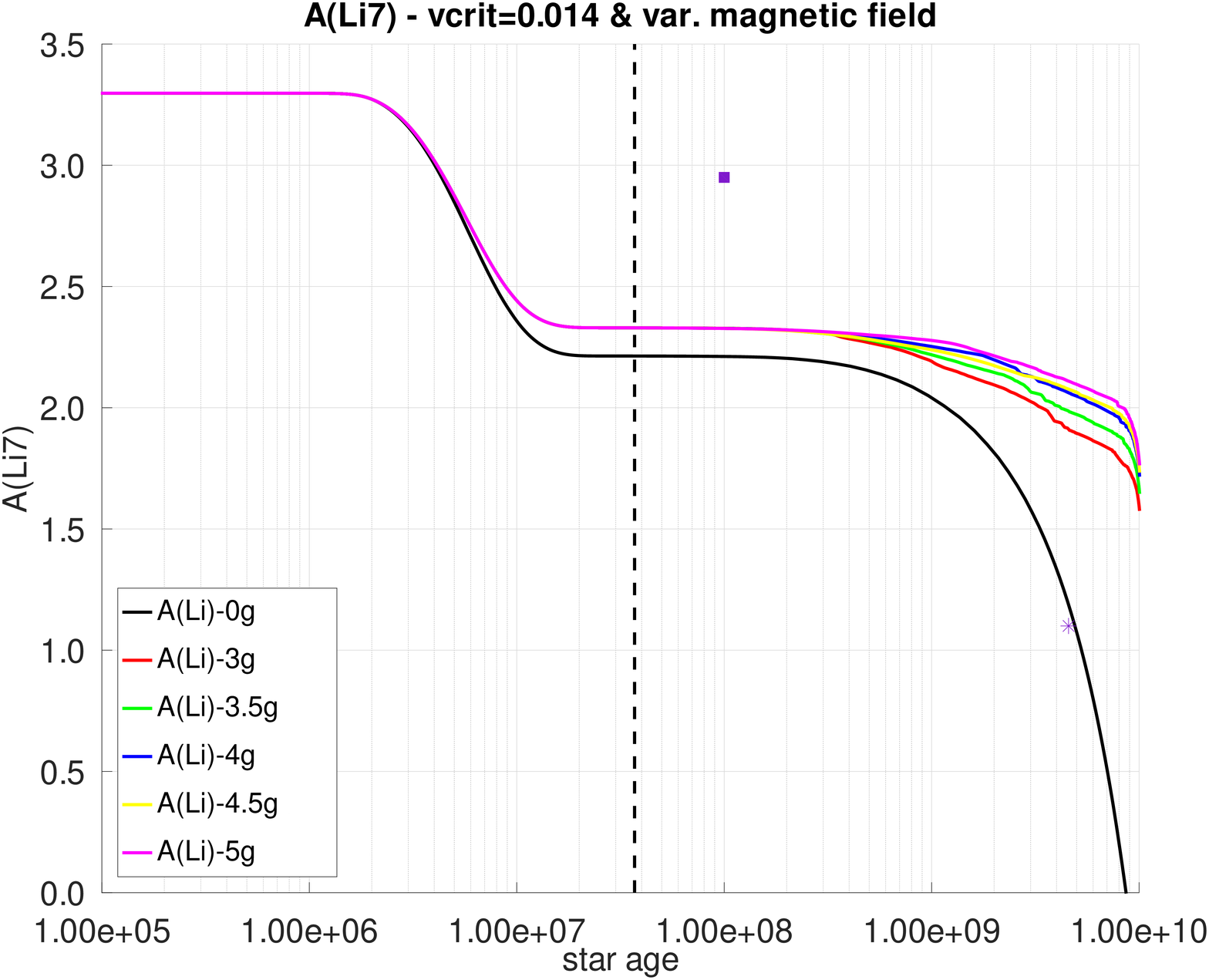}
    \label{fig:subim22}
    \end{subfigure}
    \begin{subfigure}[h]{0.47\textwidth}
    \includegraphics[trim = 25mm 10mm 15mm 10mm, clip,width=\textwidth]{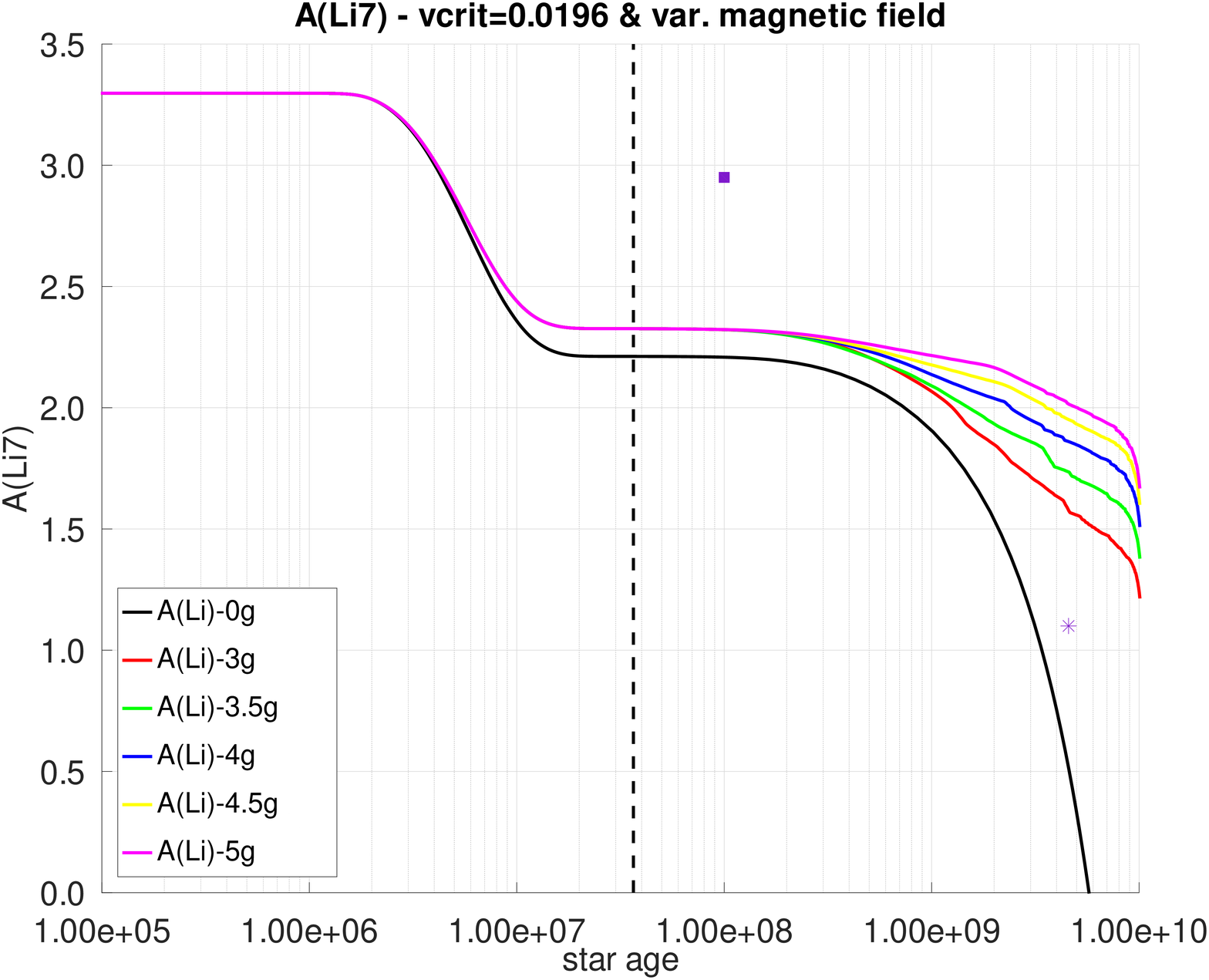}
    \label{fig:subim23}
    \end{subfigure}
    \begin{subfigure}[h]{0.47\textwidth}
    \includegraphics[trim = 25mm 10mm 15mm 10mm, clip,width=\textwidth]{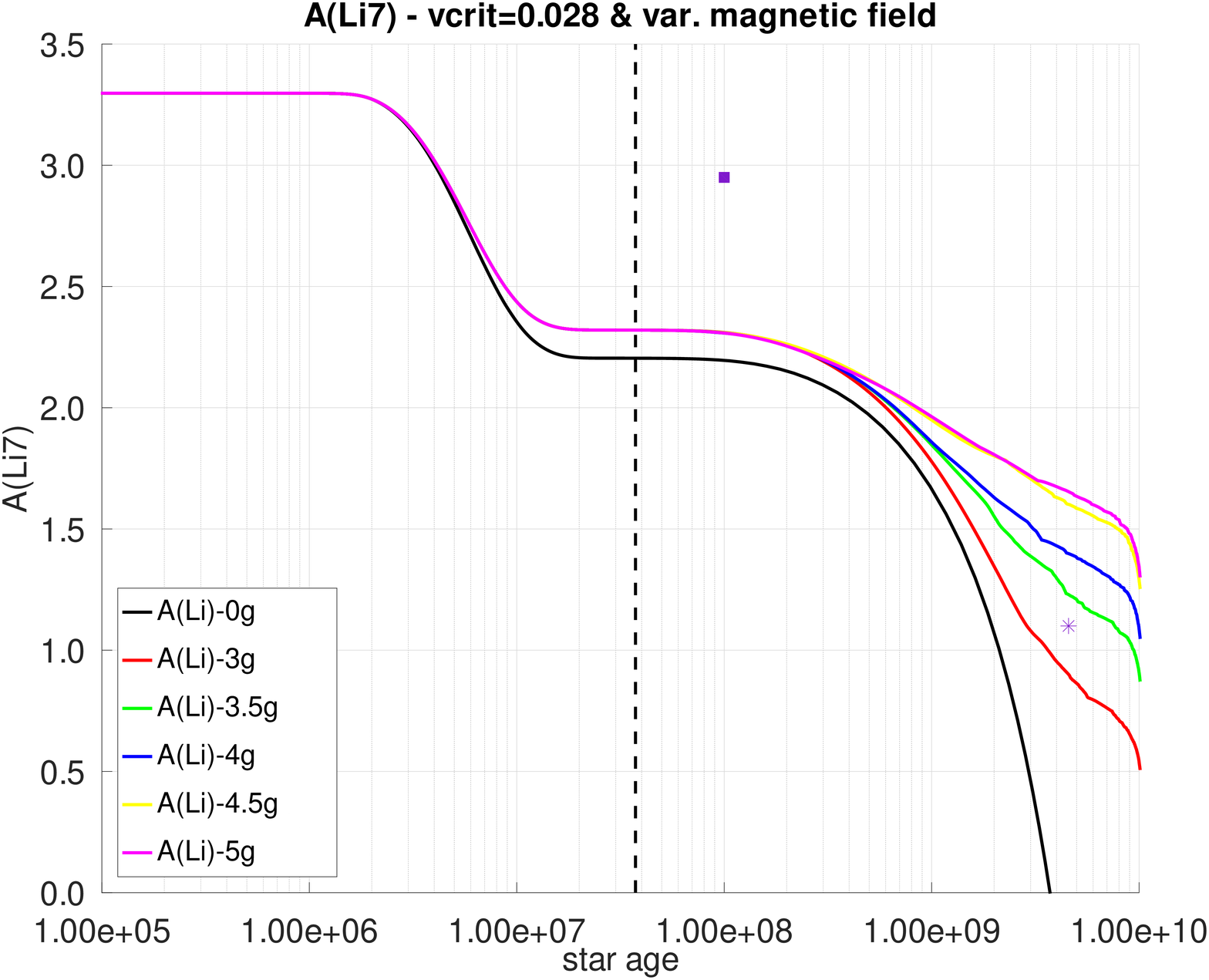}
    \label{fig:subim24}
    \end{subfigure}
    \begin{subfigure}[h]{0.47\textwidth}
    \includegraphics[trim = 25mm 10mm 15mm 10mm, clip,width=\textwidth]{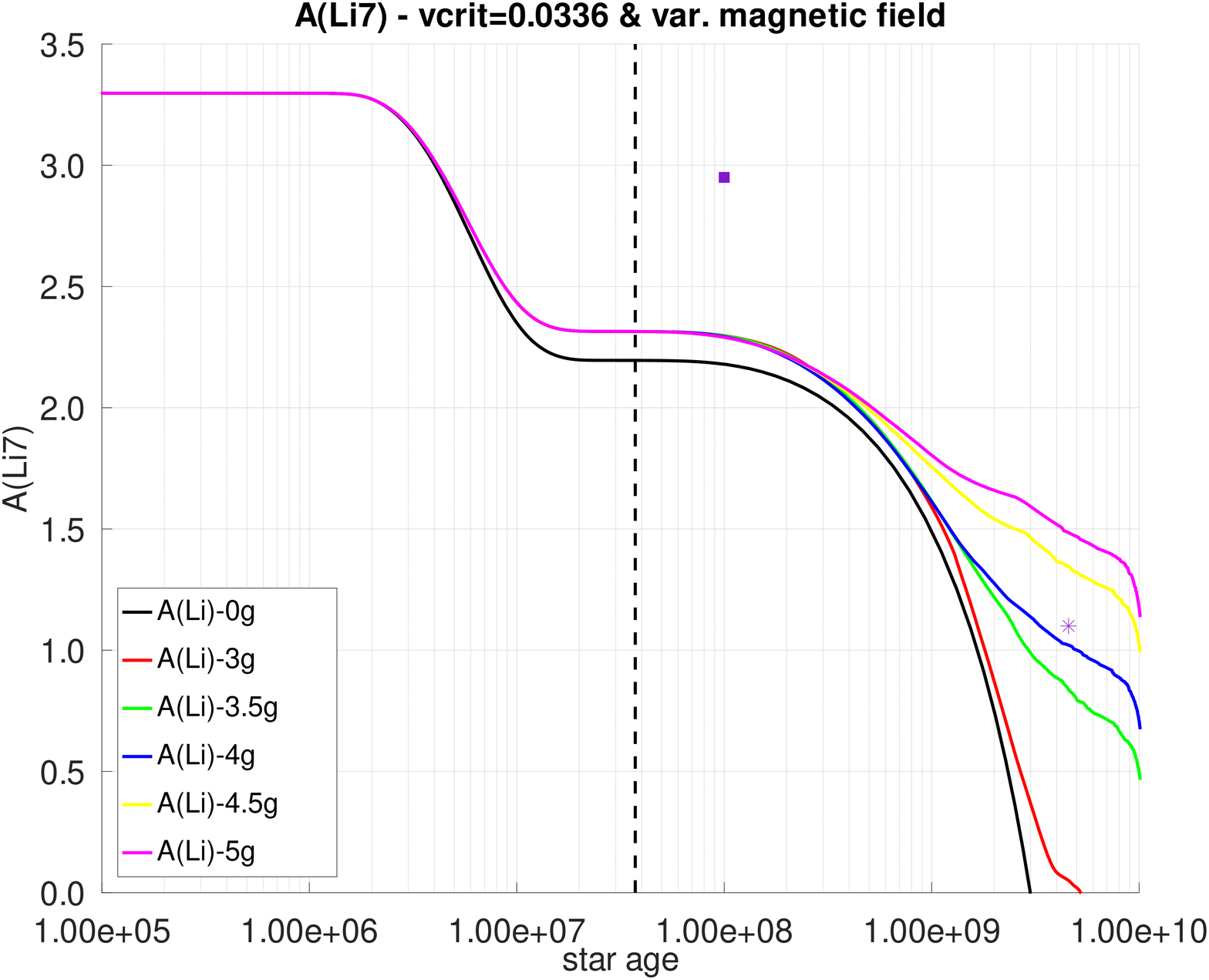}
    \label{fig:subim25}
    \end{subfigure}
    \begin{subfigure}[h]{0.47\textwidth}
    \includegraphics[width=\textwidth]{figures/blank.eps}
    \label{fig:subim26}
    \end{subfigure}
\caption{Grid showing the evolution of surface \isotope[7]{Li} abundance relative to \isotope[1]{H}, as a function of time for several 1 $\msun$ models. Each figure shows a set of models in which $\oomegac$ has been fixed and the magnetic field with intensity varies between $0.0\,\Gauss$ and $5.0\,\Gauss$, respectively. The purple star and square are surface Li abundance for the present-day Sun \citep{Asplund2009} and the Pleiades cluster \citep{Sestito2005} respectively. The dashed vertical line makes reference to the ZAMS.}
\label{fig:grid_li_var_g}
\end{figure*}

\begin{figure*}
    \centering
    \begin{subfigure}[h]{0.47\textwidth}
    \includegraphics[trim = 10mm 10mm 15mm 10mm, clip,width=\textwidth]{figures/rot_vel_var_vel_0_0g.eps}
    \label{fig:subim41}
    \end{subfigure}
    \begin{subfigure}[h]{0.47\textwidth}
    \includegraphics[trim = 10mm 10mm 15mm 10mm, clip,width=\textwidth]{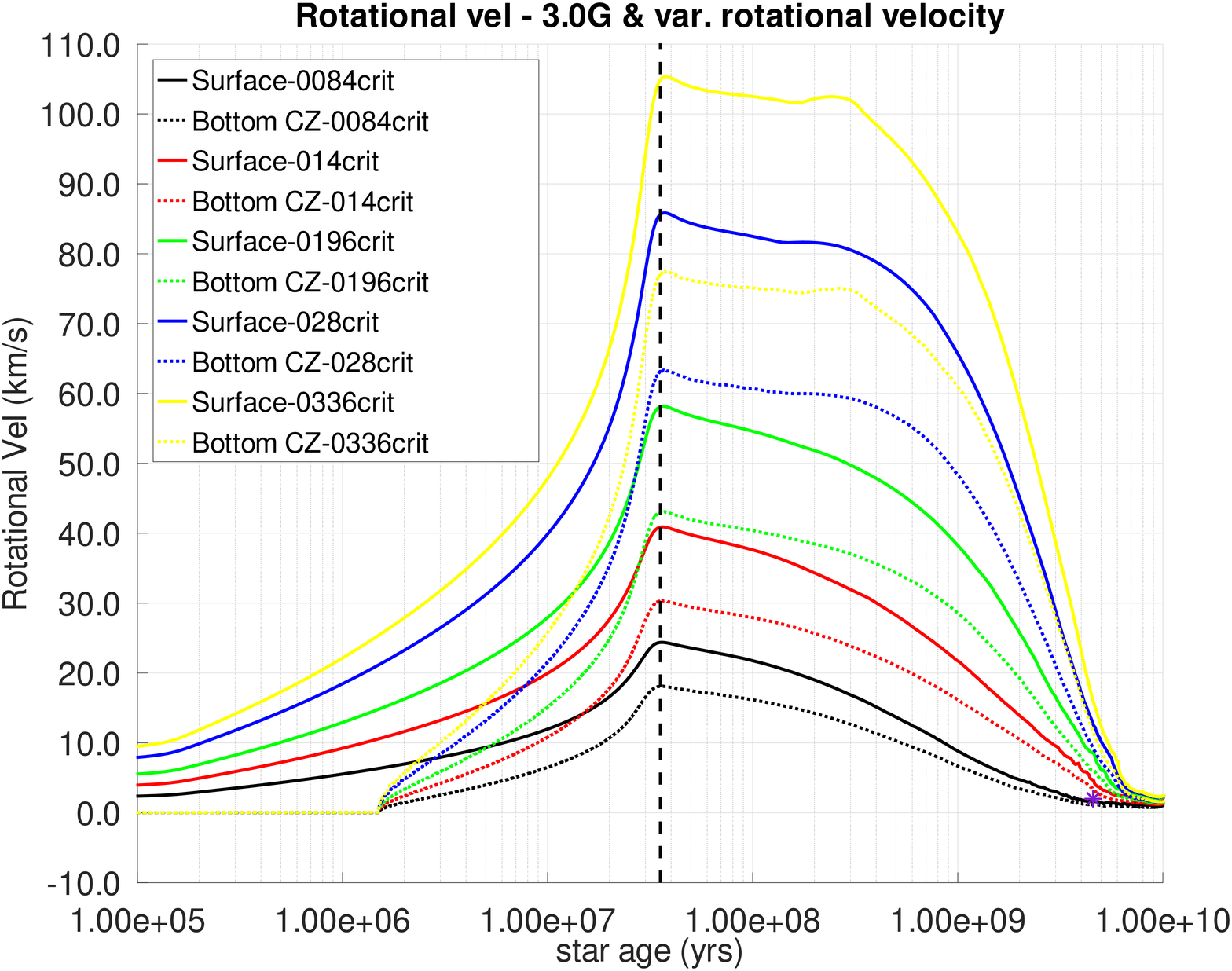}
    \label{fig:subim42}
    \end{subfigure}
    \begin{subfigure}[h]{0.47\textwidth}
    \includegraphics[trim = 10mm 10mm 15mm 10mm, clip,width=\textwidth]{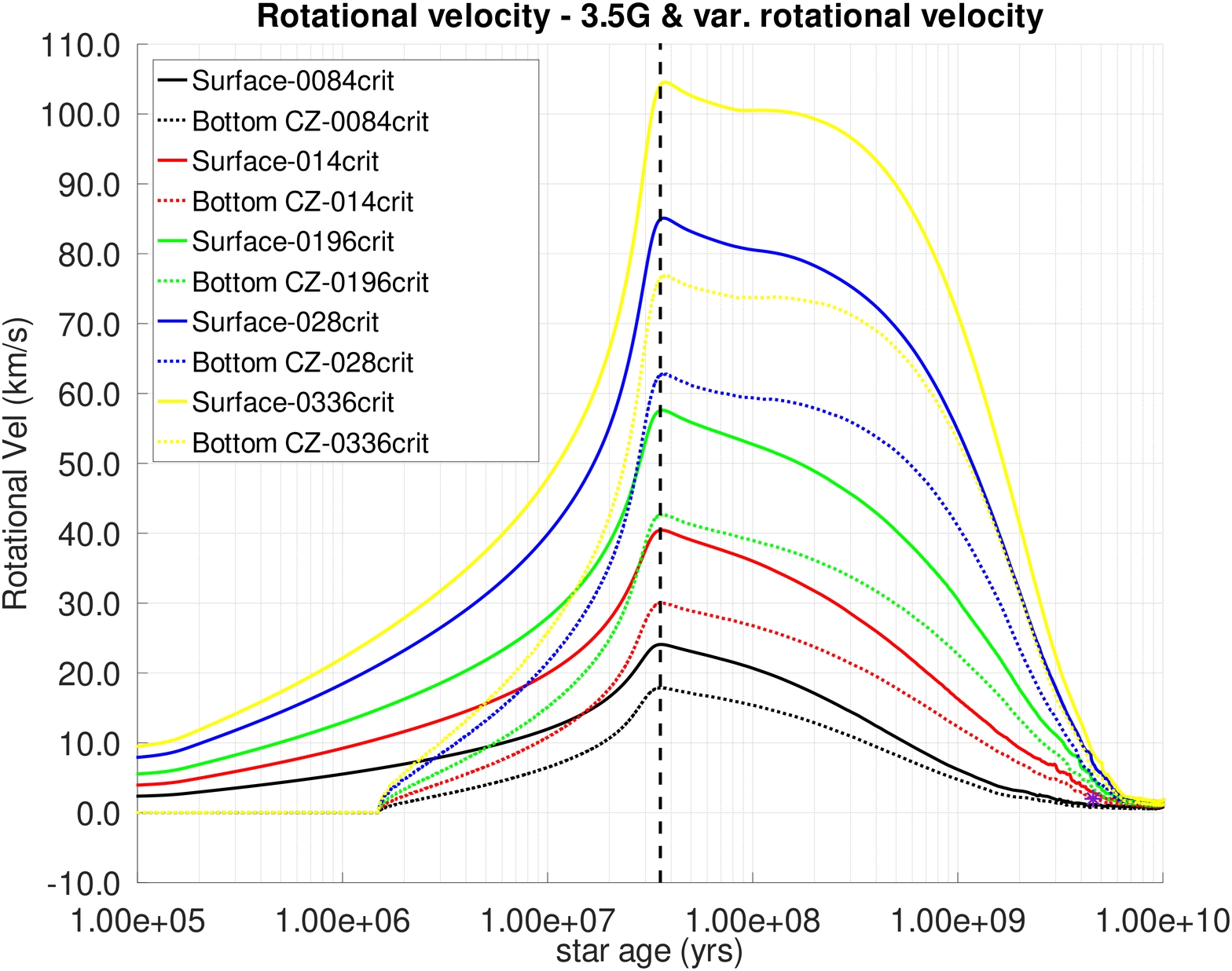}
    \label{fig:subim43}
    \end{subfigure}
    \begin{subfigure}[h]{0.47\textwidth}
    \includegraphics[trim = 10mm 10mm 15mm 10mm, clip,width=\textwidth]{figures/rot_vel_var_vel_4_0g.eps}
    \label{fig:subim44}
    \end{subfigure}
    \begin{subfigure}[h]{0.47\textwidth}
    \includegraphics[trim = 10mm 10mm 15mm 10mm, clip,width=\textwidth]{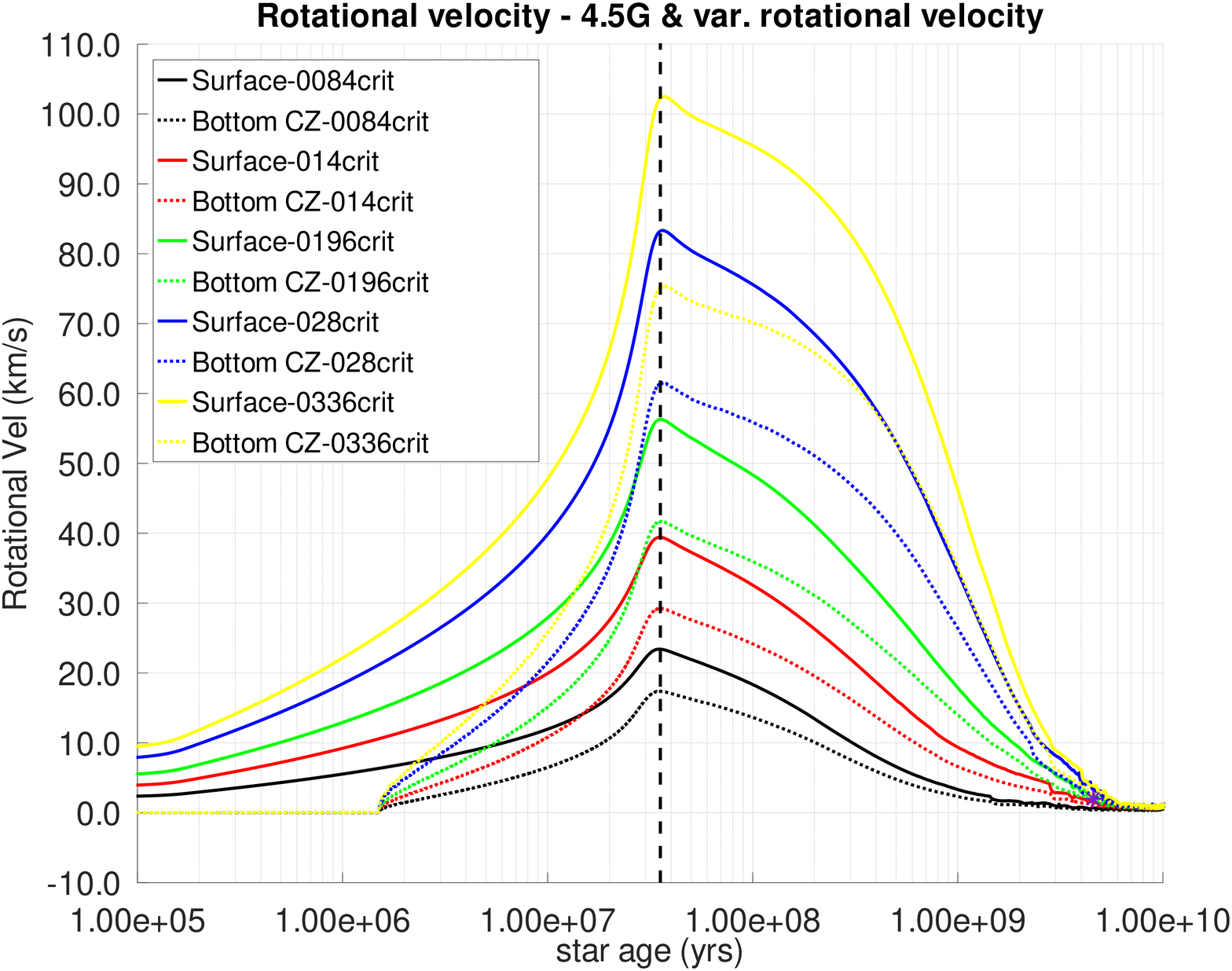}
    \label{fig:subim45}
    \end{subfigure}
    \begin{subfigure}[h]{0.47\textwidth}
    \includegraphics[trim = 10mm 10mm 15mm 10mm, clip,width=\textwidth]{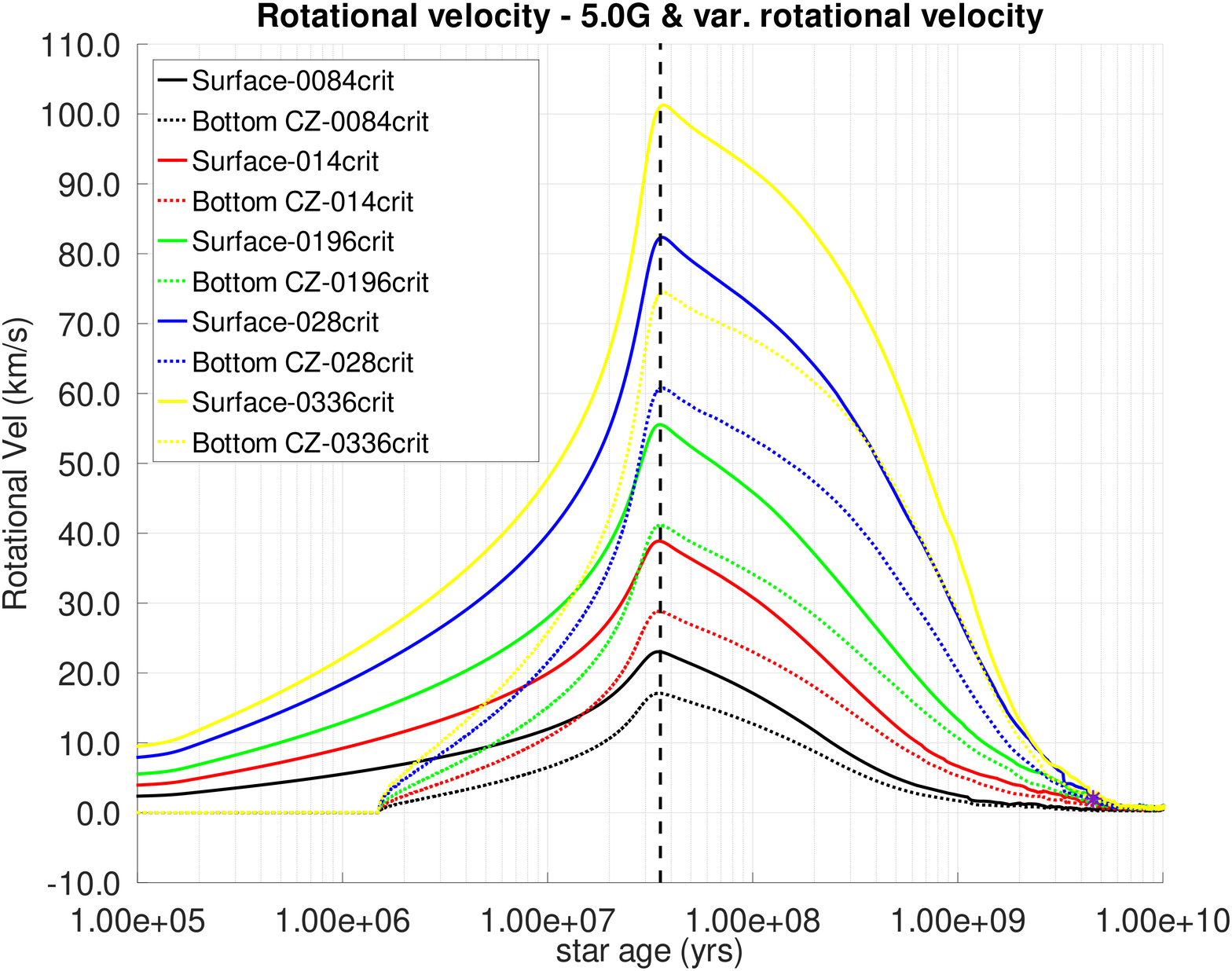}
    \label{fig:subim46}
    \end{subfigure}
\caption{Grid showing of the evolution of surface rotational velocity, as a function of time for several 1 $\msun$ models. Each figure shows a set of models in which the magnetic field with intensity has been fixed and $\oomegac$ varies between 0.0084 and 0.0336. The purple star is the surface angular velocity for the present-day Sun \citep{Gill2012}. The dashed vertical line makes reference to the ZAMS.}
\label{fig:grid_rot_vel}
\end{figure*}


\bsp	
\label{lastpage}
\end{document}

%% file: definitions.tex
\newcommand{\msun} {\mathrm{M}_\odot}
\newcommand{\mstar} {\mathrm{M}_*}

\newcommand{\rsun} {\mathrm{R}_\odot}
\newcommand{\rstar} {\mathrm{R}_*}

\newcommand{\lsun} {\mathrm{L}_\odot}
\newcommand{\llsun} {L/\mathrm{L}_\odot}

\newcommand{\kms} {\mathrm{kms}^{-1}}

\newcommand{\teff} {T_{\mathrm{eff}}}
\newcommand{\amlt} {\alpha_{\mathrm{MLT}}}

\newcommand{\oomegac}{\Omega / \Omega_{\mathrm{crit}}}
\newcommand{\Kelvin}{\mathrm{K}}
\newcommand{\Gauss}{\mathrm{G}}

\newcommand{\tli}{T_{\mathrm{Li}}}